\documentclass[usenatbib]{mnras}
\usepackage{graphicx}
\usepackage{dcolumn}
\usepackage{amssymb} 
\usepackage{amsmath}
\usepackage{comment}
\usepackage{ctable}
\usepackage{float}
\usepackage{bm}
\usepackage{float}
\usepackage{epsfig}
\usepackage{epstopdf}
\usepackage{epsf,color}
\usepackage{comment}
\usepackage{aas_macros}
\usepackage{url}
\usepackage{soul}
\usepackage{widetext}
\usepackage{color, colortbl}
\newcommand{\cmnt}[1]{}

\usepackage[normalem]{ulem}
\usepackage{color}

\usepackage{tgtermes}

\newcommand{\cii}{\ion{C}{ii}}

\newcommand{\oiii}{\ion{O}{iii}}
\newcommand{\oii}{\ion{O}{ii}}
\newcommand{\nii}{\ion{N}{ii}}

\title[\oiii\ model]{An analytic model for [\oiii] fine structure emission from high redshift galaxies}

\author[S. Yang et al.]{
Shengqi Yang,$^{1}$\thanks{E-mail:sy1823@nyu.edu}
Adam Lidz$^{2}$\\
$^{1}$Center for Cosmology and Particle Physics, Department of physics, New York University, 726 Broadway, New York, NY, 10003, U.S.A.\\
$^{2}$Department of Physics and Astronomy, University of Pennsylvania, 209 South 33rd Street, Philadelphia, PA 19104, USA
}

\date{Accepted XXX. Received YYY; in original form ZZZ}

\pubyear{2018}

\begin{document}
\label{firstpage}
\pagerange{\pageref{firstpage}--\pageref{lastpage}}
\maketitle

\begin{abstract}
Recent ALMA measurements have revealed bright [\oiii] 88 micron line emission from galaxies during the Epoch of Reionization (EoR) at redshifts as large as $z \sim 9$. We introduce an analytic model to help interpret these and other upcoming [\oiii] 88 micron measurements. Our approach sums over the emission from discrete Str$\ddot{\mathrm{o}}$mgren spheres and considers the total volume of ionized hydrogen in a galaxy of a given star-formation rate. We estimate the relative volume of doubly-ionized oxygen and ionized hydrogen and its dependence on the spectrum of ionizing photons. We then calculate the level populations of OIII ions in different fine-structure states for HII regions of specified parameters. In this simple model, a galaxy's [\oiii] 88 $\mu$m luminosity is determined by: the typical number density of free electrons in HII regions ($n_e$), the average metallicity of these regions ($Z$), the rate of hydrogen ionizing photons emitted ($Q_{\mathrm{HI}}$), and the shape of the ionizing spectrum. We cross-check our model by comparing it with detailed \textsc{CLOUDY} calculations, and find that it works to better than 15$\%$ accuracy across a broad range of parameter space. Applying our model to existing ALMA data at $z \sim 6-9$, we derive lower bounds on the gas metallicity and upper bounds on the gas density in the HII regions of these galaxies. These limits vary considerably from galaxy to galaxy, with the tightest bounds indicating $Z \gtrsim 0.5 Z_\odot$ and $n_{\mathrm{H}} \lesssim 50$ cm$^{-3}$ at $2-\sigma$ confidence.

\end{abstract}




\section{Introduction}

Observations of far-infrared fine structure emission lines have enabled spectrosopic redshift determinations for some of the highest redshift galaxies detected thus far at $z \sim 6-9$ (e.g., \cite{2015ApJ...807..180W,2016Sci...352.1559I,2018Natur.553...51M}). These measurements probe into the Epoch of Reionization (EoR) when early generations of luminous sources form, emit ultraviolet light, and gradually photo-ionize neutral hydrogen in the intergalactic medium (IGM). Fine-structure emission lines provide valuable information regarding the interstellar medium (ISM) in star-forming galaxies and the new high redshift measurements have started to probe ISM properties back into the EoR. Among other quantities, fine-structure lines may be used to constrain the gas phase metallicities, gas densities, and the spectral shape of the ionizing radiation from early stellar populations. 

Thus far, ALMA observations have yielded a handful of detections of [\cii] 158 $\mu$m and [\oiii] 88 $\mu$m emission lines at $z \sim 6-9$ (e.g., \citealt{2016ApJ...829L..11P,2017ApJ...837L..21L,2017A&A...605A..42C,2018Natur.553..178S,2018MNRAS.478.1170C,2018ApJ...854L...7C,2018Natur.557..392H,2019PASJ...71...71H,2019ApJ...874...27T,Harikane19,2019ApJ...881...63N}). An interesting result from these measurements is that the galaxies' [\oiii] luminosity is comparable or brighter than that of local galaxies with similar star-formation rates (SFRs) \citep{2014A&A...568A..62D,2018MNRAS.481L..84M}. Further, the luminosity ratio $L_{\mathrm{OIII}}/L_{\mathrm{CII}}$ appears quite high, as much as ten times larger than in local galaxies \citep{Harikane19}.\footnote{Recent work suggests, however, that the [\cii] emission is more extended than that in [\oiii]; the current observations may consequently underestimate the [\cii] luminosity and overestimate the $L_{\mathrm{OIII}}/L_{\mathrm{CII}}$ luminosity ratio \citep{2020arXiv200609402C}.} Finally, the overall success rate of detecting the [\oiii] 88 $\mu$m line via ALMA follow-up observations of photometric $z \gtrsim 6$ galaxy candidates is quite high (see e.g. \citealt{Harikane19} and references therein). 
That is, the [\oiii] 88 $\mu$m line has emerged as a valuable probe of reionization-era galaxy populations, as anticipated by \cite{2014ApJ...780L..18I}. Future ALMA observations may also detect [\oiii] 52 $\mu$m emission lines (e.g \citealt{Jones20}), while the JWST will enable the detection of rest-frame optical transitions from OIII ions into the EoR \citep{2018MNRAS.481L..84M}.

An emerging technique, complementary to the targeted ALMA observations, is line-intensity mapping \citep{2017arXiv170909066K}. In this method, one measures the spatial fluctuations in the combined emission from many individually unresolved galaxies. A number of surveys are underway to measure the [\cii] 158 $\mu$m emission from EoR era galaxies (see e.g. \citealt{2017arXiv170909066K} for a description of some of the ongoing efforts), while [\oiii] fine-structure emission lines also provide potentially interesting targets for line-intensity mapping studies \citep{2018MNRAS.481L..84M,Beane:2018dzk}. Furthermore, the SPHEREx mission will perform line-intensity mapping observations in multiple rest-frame optical [\oiii] emission lines, among others \citep{2014arXiv1412.4872D}. 

Motivated both by the recent ALMA observations and the prospects for future line-intensity mapping measurements, the goal of this paper is to develop a simple analytic model to help interpret current and future [\oiii] emission line observations. Specifically, our aim is to model correlations between the [\oiii] 88 $\mu$m luminosity of a galaxy and its SFR, and to understand how this relationship depends on the properties of the high redshift ISM. In order to test the accuracy of our model, we compare with detailed simulations using the \textsc{CLOUDY} code \citep{2017RMxAA..53..385F}. Although \textsc{CLOUDY} is valuable on its own, our analytic model helps to isolate the key physical ingredients involved in determining the strength of a galaxy's [\oiii] emission. In addition, our model may be helpful for rapidly exploring parameter space and for predicting line-intensity mapping signals, where an enormous dynamic range in length scale is relevant. In the line-intensity mapping context our work moves towards physically-motivated descriptions of line luminosity (see also e.g. \citealt{2019ApJ...887..142S}), rather than assuming empirical correlations between e.g. luminosity and SFR. The latter approach, while simple, generally involves questionable extrapolations in redshift and luminosity, and does not directly connect with high redshift galaxy properties.

As a first application of our model we use it to derive constraints on the gas-phase metallicities and densities of the $z \sim 6-9$ [\oiii] emitters observed by ALMA. Specifically, we find that the relatively large $L_{\mathrm{OIII}}/\mathrm{SFR}$ values derived previously for these galaxies imply a {\em lower bound} on their metallicities (see also \citealt{Jones20}). Further,  collisional de-excitations become important at high densities and this leads to an {\em upper bound} on the gas density in the HII regions of these galaxies. 

The plan of this paper is as follows. In \S \ref{sec:oiii_basics} we review the basic atomic structure of the OIII ion and the emission lines of interest for our study. \S \ref{sec:model} describes both our analytic modeling framework and the suite of \textsc{CLOUDY} simulations used to test it. More specifically, \S \ref{sec:oiii_lownh} gives a relationship between [\oiii] 88 $\mu$m luminosity and SFR valid at low gas density. This result is generalized in \S \ref{sec:radiative_trapping}, \S \ref{sec:broadening}, \S \ref{sec:voiii_corr}, and \S \ref{sec:pdfs}. In \S \ref{sec:ALMA_constraints} we derive constraints from current ALMA observations. Next, in \S \ref{sec:diagnostics}, we show how the [\oiii] 88 $\mu$m measurements can be combined with future observations of the 52 $\mu$ m line and optical rest frame [\oiii] lines to further constrain the electron density and temperature of high redshift HII regions (see also \citealt{Jones20}). Finally, we conclude in \S \ref{sec:conclusions} and mention possible future research directions. 

\section{OIII atomic structure and emission lines}\label{sec:oiii_basics}

\begin{figure}
    \centering
    \includegraphics[width=0.45\textwidth]{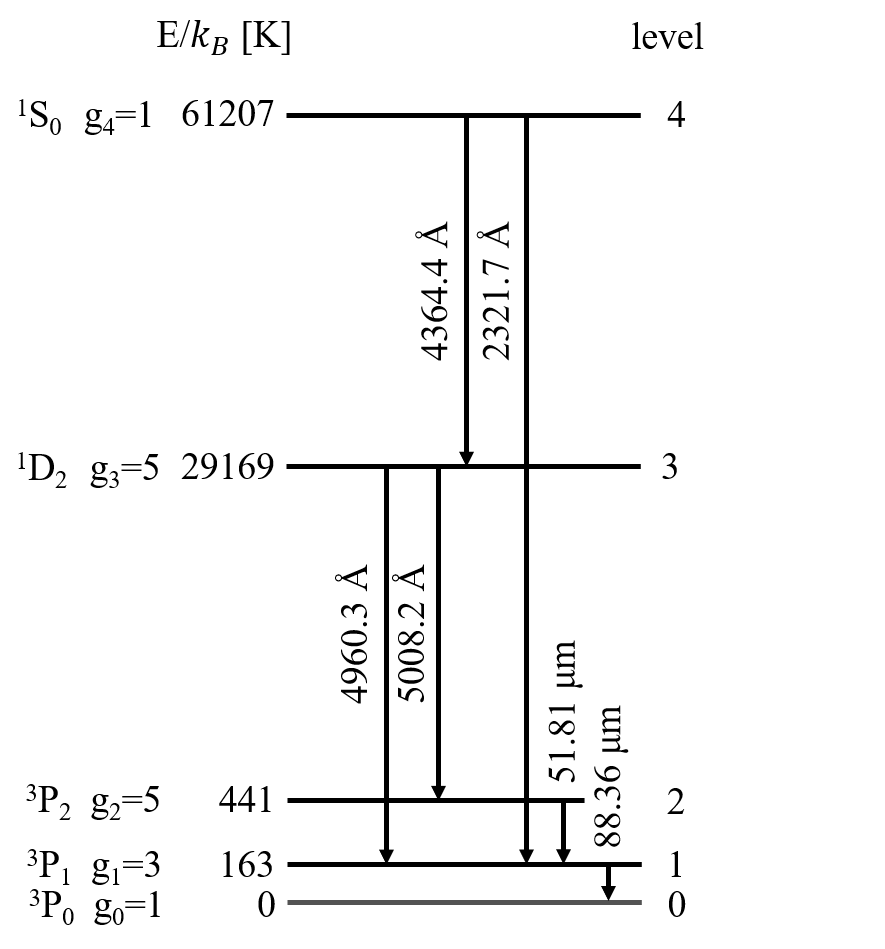}
    \caption{The bottom five energy levels of doubly ionized oxygen (OIII), including the emission lines we model in this work. The columns give (from left to right): the electronic configuration, the statistical weight, and the energy level (with energy expressed relative to the ground state in temperature units). The wavelengths of the transitions are also indicated. Adapted from \protect\cite{2011piim.book.....D} .}\label{fig:oiii_levels}
\end{figure}

Before describing our model, it is worth briefly recalling the atomic structure of the OIII ion and its main emission lines. The OIII ion consists of six electrons. In its lowest energy configuration, four electrons fill the $1s$ and $2s$ sub-shells (with two electrons in each), leaving two valence $2p$ electrons. The valence electrons may combine to form a total spin angular momentum of $S=0,1$ and a total orbital angular momentum of $L=0,1,2$. The Pauli exclusion principle allows only states with $(L,S) = (1,1); (2,0); (0,0)$. Hund's rules dictate that the lowest energy configurations have $S=1$, while the $L=2$ state lies lower than the $L=0$ one. Further, spin-orbit fine structure interactions split the lowest lying $(L,S)=(1,1)$ state into three separate levels with total (spin plus orbital) angular momentum $J=0,1,2$, ordered by increasing energy. 

The resulting energy levels and transitions are summarized in Fig \ref{fig:oiii_levels}.\footnote{The usual notation is used for the electronic configuration of each level: $^{2S+1}L_J$ with $S,P,D$ for $L=0,1,2$.}
The main line of interest for this paper is the $88.36 \mu$m transition between the $^3P_1$ and $^3P_0$ states, which has been observed by ALMA at $z \sim 6-9$. We will further briefly consider the $51.81 \mu$m fine structure transition and the rest-frame optical 5008.2 \AA\,, 4960.3 \AA\ transitions (\S \ref{sec:diagnostics}). 

\section{Model}\label{sec:model}

\begin{figure}
    \centering
    \includegraphics[width=0.45\textwidth]{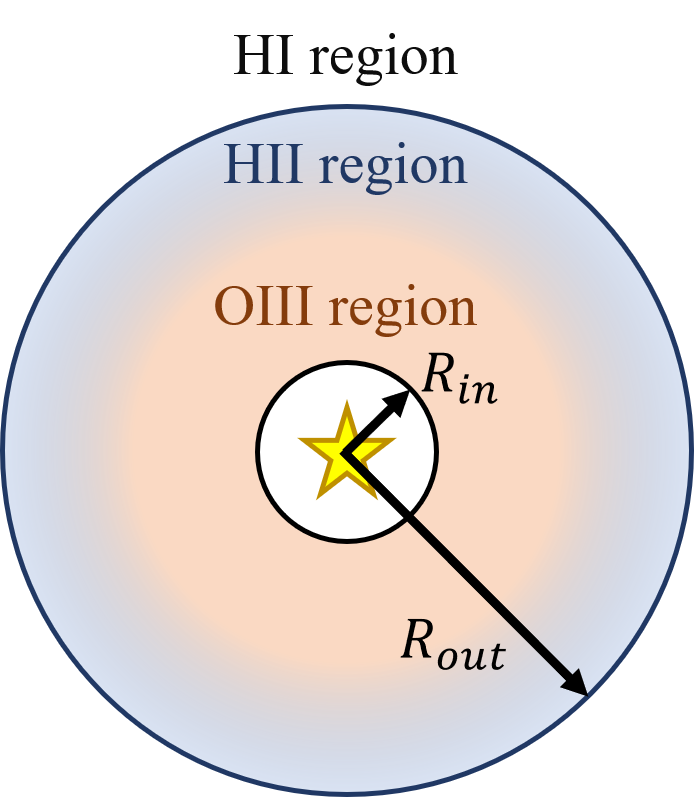}
    \caption{Simplified OIII emission geometry adopted in this work. As described in the text, the entire ionizing output from a model galaxy is concentrated in a single source at the center of a spherically symmetric region. The HII region encompasses the total volume of ionized gas in the galaxy. Our aim is to characterize the emission from the OIII gas in the interior of this region. For some parameters of interest, the OIII zone is smaller and more diffuse than the HII region as shown, although this effect is exaggerated here. }\label{fig:geometry}
\end{figure}

Our goal is to model the total [\oiii] emission from high redshift galaxies. In reality, this emission arises from a collection of discrete HII regions, distributed across the entire galaxy, each with a complex internal structure, geometry, and dynamics. Our simple model will ignore much of this complexity, as the aggregate emission from this assembly of HII regions should depend mostly on the galaxy-averaged HII region properties and the galaxy's total rate of ionizing photon production. Therefore, our model starts from the greatly simplified picture illustrated in Fig \ref{fig:geometry}. Here, the total ionizing emission from a galaxy is concentrated in a single effective source of radiation, denoted by a star in the figure, at the center of a spherical region. Further, the region is taken to be uniform in hydrogen density and metallicity. We subsequently generalize our treatment to consider an ensemble of HII regions described by probability distributions in density and metallicity (\S \ref{sec:pdfs}). A simple model like this is obviously incapable of describing the likely complex structure of the OIII emission across an individual galaxy. We believe it nevertheless properly accounts for the total galaxy-wide OIII luminosity and its dependence on typical HII region and stellar ionizing radiation parameters.
We comment on possible limitations of our treatment when necessary.\par

\subsection{\textsc{CLOUDY} simulations}\label{sec:cloudy}

As motivated in the Introduction, we compare our analytic model with detailed \textsc{CLOUDY} calculations. In order to characterize the accuracy of our model across a broad parameter space, we perform \textsc{CLOUDY} simulations with: hydrogen density between $0.5\leq\log(n_{\mathrm{H}}/\mathrm{[cm^{-3}]})\leq4$\footnote{In this work $\log$ denotes a base-10 logarithm and $\ln$ denotes a natural logarithm.}, and gas metallicities of $Z/Z_\odot=0.05$, 0.2, 1.0. We characterize the strength of the galaxy's ionizing radiation by the number of HI ionizing photons produced per second, and vary this across the range $50\leq\log(Q_{\mathrm{HI}}/\mathrm{[s^{-1}]})\leq56$. This range in ionization rate spans star-formation rates of roughly $\mathrm{SFR} \sim 10^{-3}-10^{3} M_\odot/\mathrm{yr}$ (Eq \ref{eq:q_sfr}).
The relative abundances of chemical elements in the gas cloud are scaled to the solar values \citep{2010Ap&SS.328..179G}, except for helium whose abundance is set following \cite{2004ApJS..153...75G}. As discussed in \S \ref{sec:broadening}, our simulations include turbulent velocities (with a fiducial value of $v_{\mathrm{turb}} = 100$ km/s.) We do not incorporate a turbulent pressure component.

The input stellar spectrum adopted follows either population synthesis calculations from the \textsc{starburst99} \citep{1999ApJS..123....3L} code or an effective blackbody form described below. We do not include any contributions from AGN. Our fiducial \textsc{starburst99} spectrum assumes a constant star-formation rate (SFR) and a Salpeter initial mass function (IMF) \citep{1955ApJ...121..161S}. The lower and upper cutoffs of the IMF are 1$M_\odot$ and 100$M_\odot$ respectively (although see Eq \ref{eq:q_sfr} below).  We generally assume that the stellar metallicity and the metallicity of the HII regions are identical, but we consider variations around this assumption. \cite{Steidel:2016hvv} finds that the spectra of $z \sim 2-3$ LBGs are best explained if the stellar metallicity is a factor of $\sim 5$ smaller than the gas-phase metallicity. These authors argue that this is a consequence of a super-solar oxygen to iron abundance ratio (since the stellar opacity and mass loss is largely controlled by iron), as expected for enrichment dominated by core-collapse supernovae. That is, in high redshift galaxies we should naturally expect the gas-phase metallicity to be enhanced relative to the stellar metallicity.
Finally, our fiducial model considers \textsc{starburst99} spectra for continuous SFR models with an age of 10 Myr since the ionizing spectral shape reaches an equilibrium value slightly before this duration.
Our results are therefore insensitive to this choice of age provided continuous SFR models are a good description (rather than aging starbursts), as discussed further in \S \ref{sec:voiii_corr}.

It is also interesting to consider variations in the shape of the ionizing spectrum. Recent advances in spectral synthesis modeling have emphasized the importance of binarity and rapid rotation in determining the production of ionizing photons from massive stars (e.g. \citealt{Eldridge12,Stanway20}). These effects tend to make the ionizing spectral shape harder than otherwise expected. Here we defer a full treatment to future work and follow \cite{Steidel:2014iea} in simply approximating the ionizing radiation spectral shape between 1 and 4 Rydbergs by a blackbody form, characterized by an effective temperature (in units of $10^4$ K), $T_{4,\mathrm{eff}}$. This description is intended to avoid a full suite of population synthesis models, while nevertheless spanning a range of plausible spectral shapes. 

We run \textsc{CLOUDY} in its spherical geometry mode, with an inner radius of $0.01$ times the Str$\ddot{\mathrm{o}}$mgren radius of the HII region. We assume an ionization-bounded HII region, and ignore the impact of magnetic fields, dust grains, and external radiation fields. We comment on the possible impact of dust grains in Section \ref{sec:conclusions}. The \textsc{CLOUDY} calculations are halted when the electron density drops to half of the input hydrogen density. 

\subsection{Analytic model setup}

Similar to the \textsc{CLOUDY} simulations, our analytic model is parameterized by: $n_\mathrm{H}$, $Z/Z_\odot$, and an SFR.  In \S \ref{sec:pdfs} we further allow for variations in $n_{\mathrm{H}}$ and $Z/Z_\odot$ characterized by lognormal distributions. The shape of the stellar spectrum also plays an important role. Our default assumptions regarding this spectrum are identical to those in the \textsc{CLOUDY} simulations discussed above. 

In terms of the atomic data in our modeling, we use results from the following references throughout this paper. We employ the fits to the photoionization cross-sections from \cite{1996ApJ...465..487V}, the Einstein A coefficients given by \cite{1973OPurA...5..192G} Table 1, the case B recombination rate of HII and HeII from \cite{2011piim.book.....D} Table 14.1 and Table 14.7, the OIII collisional de-excitation rate given by \cite{2011piim.book.....D} Section 2.3 and Table F.2, the OIII recombination rates from \cite{2006agna.book.....O} Table A5.1, and the charge-exchange rates from \cite{2006agna.book.....O} Table A5.4.

The Str$\ddot{\mathrm{o}}$mgren radius provides a key description of the ionized hydrogen regions in our model. It is important to note that the volume discussed here indicates the total volume of ionized hydrogen gas in the galaxy (as in Fig \ref{fig:geometry} and discussed previously), rather than that around an individual star or star cluster. We assume that a fraction $f_\mathrm{esc}$ of the hydrogen ionizing photons escape the galaxy and make it into the intergalactic medium (IGM), while the rest are absorbed within the galaxy. Unless stated otherwise, we assume 
$1 - f_{\mathrm{esc}} = 1$ since empirically -- and in many theoretical models --  most galaxies have small escape fractions \citep{2000ApJ...531..846D,2000ApJ...545...86W,2010ApJ...710.1239R,2008ApJ...672..765G,2009ApJ...693..984W}. It is straightforward to rescale our results for the case of larger escape fractions. 
Assuming $1 - f_{\mathrm{esc}}=1$ the Str$\ddot{\mathrm{o}}$mgren radius of the HII region, R$_{\mathrm{out}}$, is determined by\footnote{This assumes photo-ionization equilibrium, which is justified because the recombination timescale, $t_{\mathrm{rec}}=1/(\alpha_B n_e) = 1.2 \times 10^3 ~\ {\mathrm{yrs}} ~\ (n_e/100 \mathrm{cm}^{-3})^{-1}$, is much shorter than the age of O stars and the time over which the rate of ionizing photon production evolves.}
\begin{equation}\label{eq:1}
    (1-f_\mathrm{esc}) Q_{\mathrm{HI}}=\dfrac{4\pi}{3}R_{\mathrm{out}}^3 \alpha_{B,\mathrm{HII}}(T)n_{\mathrm{H}}n_e= V_{\mathrm{HII}}\alpha_{B,\mathrm{HII}}(T)n_{\mathrm{H}}n_e\,,
\end{equation}
where $n_{\mathrm{H}}$ is the hydrogen density, $n_e$ is the electron density, $V_{\mathrm{HII}}$ is the volume of HII region (again, here and throughout this refers to the total volume of ionized hydrogen gas in the galaxy), and $\alpha_{B,\mathrm{HII}}$ is the case B recombination rate of hydrogen \citep{1987MNRAS.224..801H}.\par

Our model and the \textsc{CLOUDY} calculations assume thermal equilibrium applies, with heating and cooling process balancing in a steady-state description. We approximate the temperature, $T$, by its volume-averaged value. In practice the temperature is difficult to predict analytically, and is fairly close to $T_4=1$ across the parameter space of interest. Therefore, the temperature is the one aspect of our model that we simply calibrate empirically to \textsc{CLOUDY} simulations. In particular, the following fit on $T_4=T/(10^4$ [K]) is accurate to within 15\% fractional error across the full range of \textsc{CLOUDY} models (spanning $n_{\mathrm{H}}$, $Z$, and $Q_{\mathrm{HI}}$) mentioned previously\footnote{This fit assumes the 10 Myr continuous SFR \textsc{starburst99} model. Since the spectral shape reaches an equilibrium value at slightly smaller ages, it should be a good description for most continuous SFR models. The fit should, however, be modified for the case of aging starbursts.}:
\begin{equation}\label{eq:3}
\begin{split}
    T_4&=10^{-0.16(\log \frac{Z}{Z_\odot})^2-0.6\log \frac{Z}{Z_\odot}-0.18}\times\dfrac{\log \frac{Q_{\mathrm{HI}}}{[\mathrm{s^{-1}}]}}{54}\\
    &\times\left((0.1\left(\frac{Z}{Z_\odot}\right)^{0.6}\log \left(\frac{n_{\mathrm{HI}}}{[\mathrm{cm^{-3}}]}\right)+0.9\right)\,,
\end{split}
\end{equation}\par
It is further interesting to relate the ionizing photon production rate, $Q_{\mathrm{HI}}$, to the instantaneous star-formation rate, SFR, across the galaxy. A fitting formula to population synthesis calculations, valid for constant SFR models with age larger than 6 Myr, from \cite{2003A&A...397..527S} provides a convenient relationship:
\begin{equation}\label{eq:q_sfr}
\begin{split}
    &\log\left(\dfrac{Q_{\mathrm{HI}}/\ \mathrm{[s^{-1}]}}{\mathrm{SFR}/\ [M_\odot/\mathrm{yr}]}\right)\\
    =&-0.0029\times\left(\log\left(\dfrac{Z}{Z_\odot}\right)+7.3\right)^{2.5}+53.81-\log (2.55)\,.
\end{split}
\end{equation}
This equation assumes the same Salpeter IMF discussed previously, except with a correction factor of 2.55 introduced to account for stars with mass between 0.1 and 1 $M_\odot$ \citep{2010A&A...523A..64R}.

\subsection{OIII emission in the low $n_{\mathrm{H}}$ limit}\label{sec:oiii_lownh}

We now turn to consider OIII $88\mu$m fine structure emission in our model, handling first the case of the low $n_{\mathrm{H}}$ limit. In most of this paper, we will work in the three-level atom approximation accounting for the level populations in only the $^3\mathrm{P}_{0,1,2}$ states. This is a very good approximation owing to the large energy gap between these levels and the higher energy states (see Fig \ref{fig:oiii_levels}.) The abundance of OIII ions in the $^3\mathrm{P}_{0,1,2}$ states are denoted by $n_{0,1,2}$ respectively.

For starters, we assume that our model galaxy is optically thin to OIII emission and so ignore the possibility that $88\mu$m photons are absorbed elsewhere in the galaxy that produces them. Furthermore, we ignore the impact of any external radiation fields on the level populations.\footnote{This should be an excellent approximation. Assuming that the external radiation is dominated by the cosmic microwave background (CMB), the photon occupation number at $z=7$ and 88 $\mu$m is $n_\gamma = 5.4 \times 10^{-4}$. The rate of $^3\mathrm{P}_0\, \rightarrow\, ^3 \mathrm{P}_1$ excitations after absorbing a CMB photon is $g_1 n_\gamma A_{10}/g_0 = 4.4 \times 10^{-8}\, \mathrm{s}^{-1}$. This is two orders of magnitude smaller than the collisional excitation rate for $n_e = 100\, \mathrm{cm}^{-3}$ (for $T_4=1$), and becomes comparable to the rate of collisional excitations only at $n_e = 1\, \mathrm{cm}^{-3}$. We therefore neglect the impact of the CMB throughout since it would only be relevant at very low densities (which seem unlikely). Note that the low photon occupation number also implies that stimulated emission off of the CMB is negligible. The CMB is even less important at redshifts below the $z=7$ case described explicitly here.} In this case, the level populations are determined solely by spontaneous decays and collisional excitations/de-excitations induced by collisions with free electrons. The rate of spontaneous decays between an upper level, $u$, and a lower level, $l$, is described by the Einstein A coefficient, $A_{ul}$. The collisional excitation (de-excitation) rate is given by $n_e k_{lu}$ ($n_e k_{ul})$. Each of these rates has units of s$^{-1}$. 

Since the timescales involved in the atomic transitions here is short compared to those related to star-formation and galaxy evolution, a steady-state solution applies. In the steady-state the rate of OIII ions leaving a given level exactly balances the rate at which the same level is populated. Applying this balance to the $^3\mathrm{P}_0$ level (with abundance $n_0$):
\begin{equation}\label{eq:6}
    n_0n_e(k_{01}+k_{02})=n_1(A_{10}+n_ek_{10})\Longrightarrow n_1=\dfrac{n_0n_e(k_{01}+k_{02})}{A_{10}+n_ek_{10}}\,.
\end{equation}
The left hand side describes the rate at which OIII ions are collisionally excited out of the $^3 \mathrm{P}_0$ state. The right hand side describes spontaneous decays and collisional de-excitations, which each populate the $^3\mathrm{P}_0$ level from ions initially in the $^3\mathrm{P}_1$ state (with population $n_1$).\footnote{Note that this equation neglects spontaneous decays and collisional de-exciations from the $^3\mathrm{P}_2$ state, which in principle could populate $^3\mathrm{P}_0$ as well. The spontaneous decay between these two states proceeds at a negligibly small rate, while collisional de-excitations may be ignored in the low density limit adopted here.} Further, we take the low density limit in which collisional de-excitations are negligible compared to spontaneous decays. That is, $n_e k_{10} \ll A_{10}$. For reference, the density at which spontaneous decays and collisional de-excitations make equal contribution defines the critical density, $n_\mathrm{crit} = A_{10}/k_{10} = 1732$ cm$^{-3}$ at $T_4=1$. For densities much less than the critical density, Eq \ref{eq:6} simplifies to
\begin{equation}\label{eq:n1}
    n_1\approx\dfrac{n_0n_e(k_{01}+k_{02})}{A_{10}}.
\end{equation}

Further, the total [\oiii] emission may be determined from the level population, $n_1$, as an integral over the volume containing doubly-ionized oxygen:
\begin{equation}\label{eq:l10_low}
\begin{split}
    L_{10}&=\int4\pi r^2drn_1A_{10}h\nu_{10}=n_1A_{10}h\nu_{10}V_{\mathrm{OIII}}\\
    &=n_1A_{10}h\nu_{10}V_{\mathrm{OIII}}\dfrac{Q_{\mathrm{HI}}}{V_{\mathrm{HII}}\alpha_{B,\mathrm{HII}} n_{\mathrm{H}}n_e}.
\end{split}
\end{equation}
Here we have assumed the gas is optically thin to [\oiii] emission. Note that we assume that hydrogen is mostly ionized within the HII region and that the last factor is unity with $Q_\mathrm{HI}/(\alpha_{B,\mathrm{HII}} n_\mathrm{H}n_e) = V_{\mathrm{HII}}$. 
We can then plug Eq \ref{eq:n1} into Eq \ref{eq:l10_low}, obtaining:
\begin{equation}\label{eq:l10_lowb}
\begin{split}
    L_{10}&=\dfrac{n_0}{n_{\mathrm{H}}}(k_{01}+k_{02})h\nu_{10}\dfrac{Q_{\mathrm{HI}}}{\alpha_{\mathrm{B,HII}}}\dfrac{V_{\mathrm{OIII}}}{V_{\mathrm{HII}}}\\
    &\approx\left(\dfrac{n_{\mathrm{O}}}{n_{\mathrm{H}}}\right)_\odot\dfrac{Z}{Z_\odot}(k_{01}+k_{02})h\nu_{10}\dfrac{Q_{\mathrm{HI}}}{\alpha_{\mathrm{B,HII}}}\dfrac{V_{\mathrm{OIII}}}{V_{\mathrm{HII}}}\,.
\end{split}
\end{equation}
Here we assume that the oxygen is mostly doubly-ionized within the volume $V_{\mathrm{OIII}}$ and the doubly-ionized oxygen are mostly at the ground $^3\mathrm{P}_0$ state, so that $n_0 = n_{\mathrm{O}}$ where $n_{\mathrm{O}}$ is the total oxygen abundance. Alternatively, as discussed further in \S \ref{sec:voiii_corr}, one can think of $V_{\mathrm{OIII}}/V_{\mathrm{HII}}$ as the OIII fraction averaged across the HII region. This equation supposes that all of the doubly-ionized oxygen atoms are in the ground state, which is an excellent approximation at low densities and for plausible HII region temperatures.

In the case of a hard stellar radiation spectrum with a sufficient number of photons above the OII ionization threshold at an energy of 35.12 eV (see \S \ref{sec:voiii_corr}), one can approximate $V_{\mathrm{OIII}} \approx V_{\mathrm{HII}}$. In this case, Eq \ref{eq:l10_lowb} can be simplified as:
\begin{equation}\label{eq:l10low_qhi}
    L_{10}=\left(\dfrac{n_{\mathrm{O}}}{n_{\mathrm{H}}}\right)_\odot\dfrac{Z}{Z_\odot}(k_{01}+k_{02})h\nu_{10}\dfrac{Q_{\mathrm{HI}}}{\alpha_{\mathrm{B,HII}}}\,,
\end{equation}
which is independent of hydrogen density. To summarize, this equation applies for densities much less than the critical density ($n_{\mathrm{crit}} = 1.7 \times 10^3$ cm$^{-3}$), assumes $V_{\mathrm{OIII}} \sim V_{\mathrm{HII}}$, and ignores self-absorption of [\oiii] 88$\mu$m photons, and external radiation fields.
Similar equations have been obtained in the context of [\nii] emission lines in previous work (e.g.  \citealt{2019ApJ...887..142S,2016ApJ...826..175H,1997ApJ...476..144M}).

Finally, we can connect the ionizing photon output to the SFR using Eq \ref{eq:q_sfr}. This gives:
\begin{equation}\label{eq:l10low_sfr}
   \dfrac{L_{10}}{L_\odot}=10^{8.27-0.0029(7.3+\log(Z/Z_\odot))^{2.5}}\dfrac{Z}{Z_\odot}\dfrac{\mathrm{SFR}}{M_\odot/\mathrm{yr}}\,,
\end{equation}
for $T=10^4$ K and $(n_{\mathrm{O}}/n_{\mathrm{H}})_\odot=10^{-3.31}$. When applicable, this equation implies that the [\oiii] 88 $\mu$m emission from a galaxy is directly proportional to its SFR. In addition, the luminosity varies linearly with the metallicity of the HII region, with a further dependence on stellar metallicity arising through the $Q_{\mathrm{HI}}-\mathrm{SFR}$ relationship.

\begin{figure}
    \centering
    \includegraphics[width=0.45\textwidth]{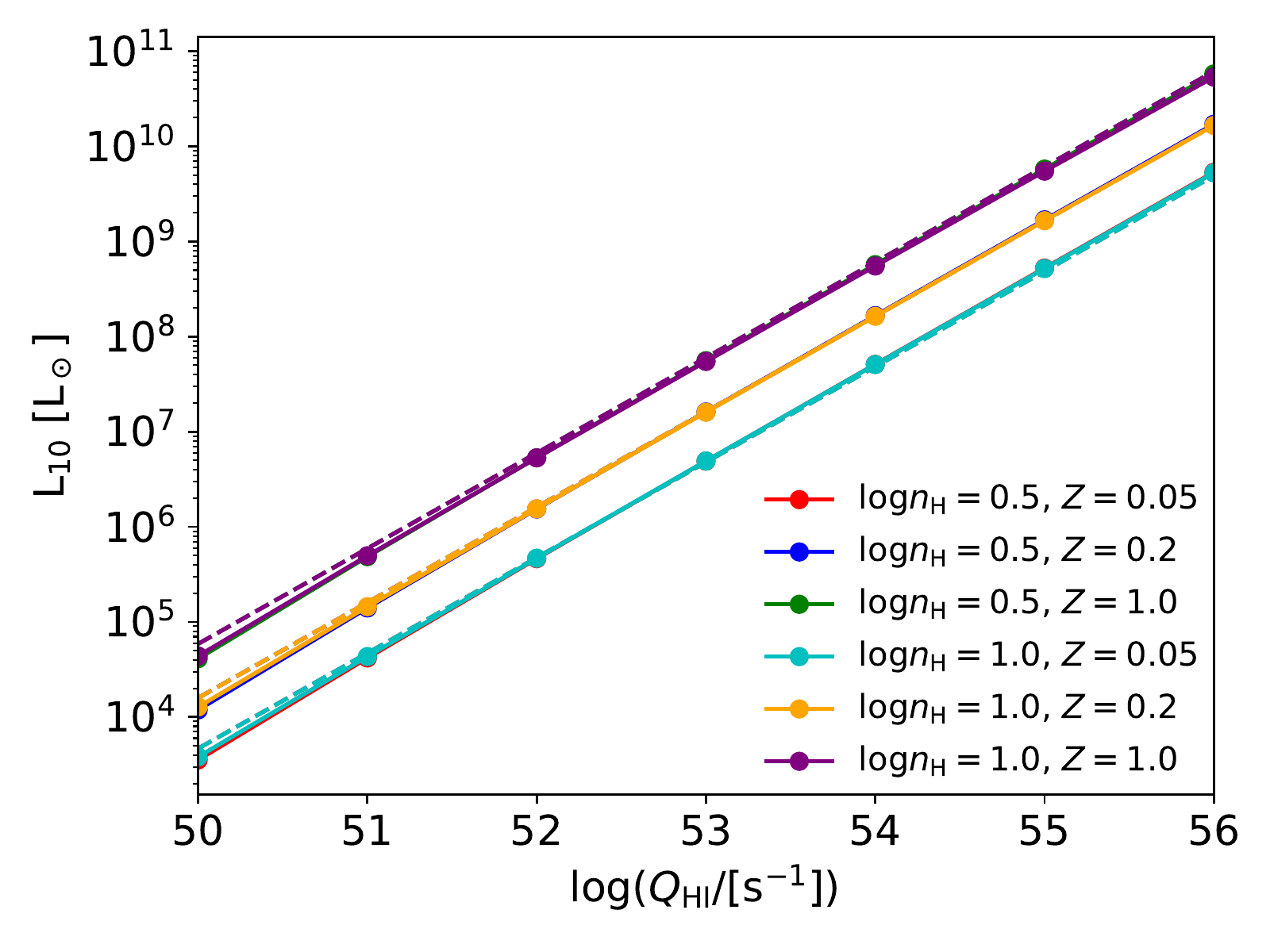}\\
    \includegraphics[width=0.45\textwidth]{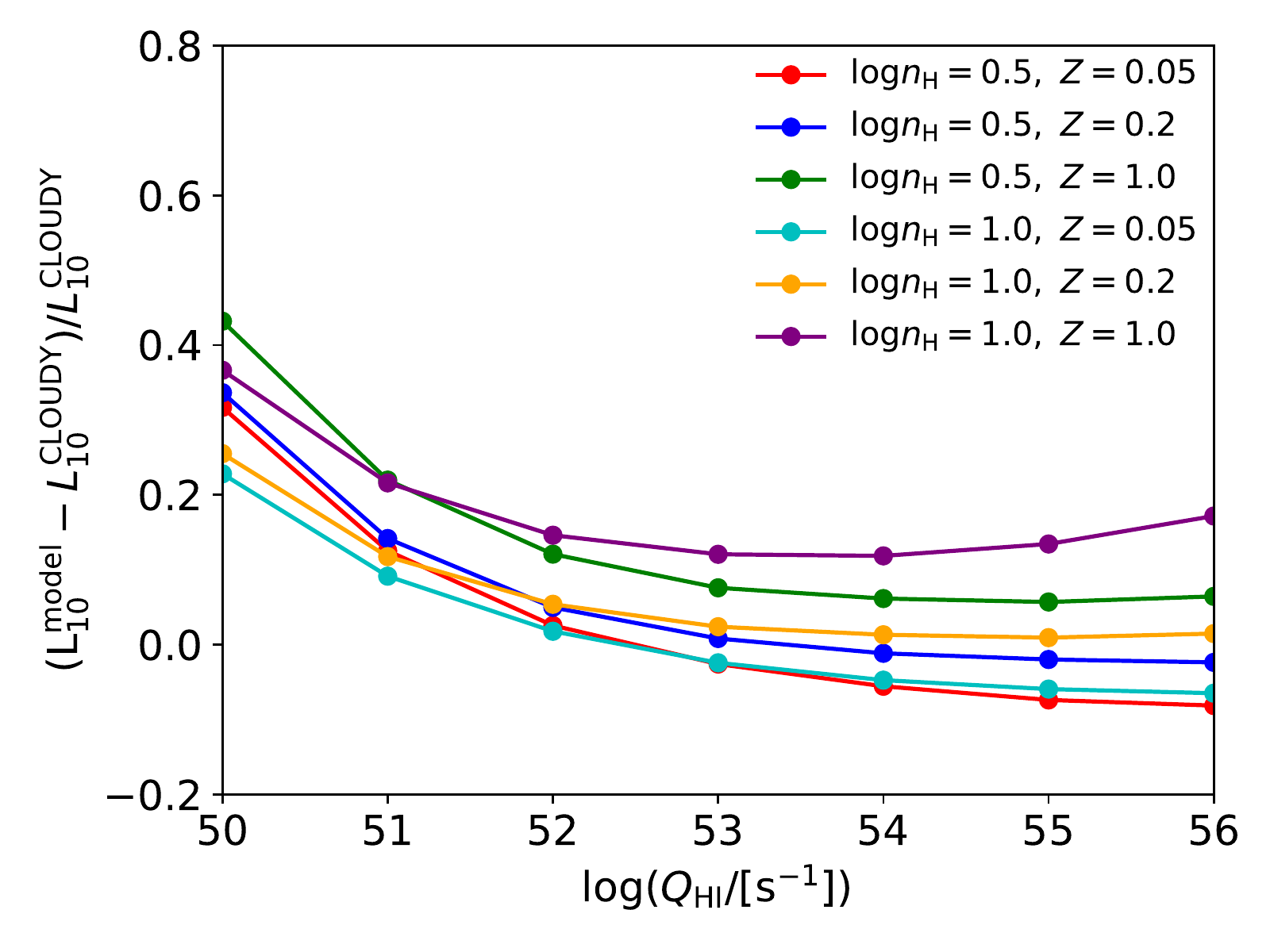}
    \caption{Comparison between the [\oiii] emission predicted by Eq \ref{eq:l10low_qhi} and \textsc{CLOUDY} simulations at low density. {\em Top panel:} Solid lines show \textsc{CLOUDY} results, while the dashed lines show the analytic model computed in the low density limit. The legend specifies the hydrogen density, $\log (\mathrm{n_{\mathrm{H}}/[cm^{-3}]})$, while $Z$ is the gas metallicity in solar units, $Z/Z_\odot$. The stellar metallicity is assumed to match the gas metallicity here. 
    {\em Bottom panel:} Fractional difference between the analytic model and the \textsc{CLOUDY} calculations.}\label{fig:l10low_sfr_accuracy}
\end{figure}

Fig \ref{fig:l10low_sfr_accuracy} compares the $L_{10}-Q_{\mathrm{HI}}$ relation of Eq \ref{eq:l10low_qhi} with \textsc{CLOUDY} simulations in the low density limit. At low densities and metallicities, the equation agrees to within 20\% fractional error and is still more accurate for much of the parameter space shown. At low $Q_{\mathrm{HI}}$ the equation slightly overpredicts the luminosity found with \textsc{CLOUDY}. This owes to the $V_{\mathrm{OIII}} \sim V_{\mathrm{HII}}$ approximation, which we will refine in \S \ref{sec:voiii_corr}. At large metallicity $Z=Z_\odot$, our model is less accurate but this case is of less interest for our aim of modeling high redshift galaxies. The model is less successful at high metallicity because the helium abundance in our model grows with metallicity \citep{2004ApJS..153...75G}, in order to account for the conversion of hydrogen into helium and heavier elements by stars. The larger helium abundance, in turn, makes the $V_{\mathrm{OIII}} \sim V_{\mathrm{HII}}$ approximation less accurate.

\subsection{Arbitrary density and the escape probability approximation}\label{sec:radiative_trapping}

Next we extend our model to arbitrary densities, accounting for collisional de-excitations in a full treatment of the three-level atom. At high densities, $L_{10}/\mathrm{SFR}$ will be suppressed relative to Eq \ref{eq:l10low_sfr} since collisional de-excitations will lead to $^3\mathrm{P}_1 \rightarrow ^3\mathrm{P}_0$ transitions without photon emission. We will also allow for the possibility that [\oiii] 88 $\mu$m photons emitted are re-absorbed somewhere in the HII region. We will model this using the escape probability approximation (e.g. \citealt{2011piim.book.....D}). 

In the escape probability approximation adopted here, the HII region is treated as static, homogeneous, and spherically symmetric so that the excitation of OIII ions across the cloud by trapped photons is uniform. The probability that photons escape the cloud, $\beta$, is then approximated by its angle and frequency averaged value, $\langle{\beta\rangle}$, assuming a Gaussian velocity distribution. The escape probability is well-approximated by \citep{2011piim.book.....D}:
\begin{equation}
    \langle \beta\rangle=\dfrac{1}{1+0.5\tau}\,.
\end{equation}
where the optical depth, $\tau$, is the line-center optical depth for a ray from the center of the region to its edge. We account for self-absorption and stimulated emission in each of the $^3\mathrm{P}_1 \rightarrow ^3\mathrm{P}_0$ and $^3\mathrm{P}_2 \rightarrow ^3\mathrm{P}_1$ transitions, while radiative trapping in the $^3\mathrm{P}_2 \rightarrow ^3\mathrm{P}_0$ line is negligible given its much smaller Einstein-A coefficient. In the escape probability approximation, radiative decay rates are reduced by a factor of $\langle{\beta\rangle}$ with $A_{ul} \rightarrow \langle{\beta\rangle} A_{ul}$. 

In the three-level atom the steady-state level populations are then determined by the following set of equations \citep{2011piim.book.....D}:
\begin{equation}\label{eq:level_ratios}
    \begin{split}
        R_1=\dfrac{n_1}{n_0}&=\dfrac{R_{01}R_{20}+R_{01}R_{21}+R_{21}R_{02}}{R_{10}R_{20}+R_{10}R_{21}+R_{12}R_{20}}\,,\\
        R_2=\dfrac{n_2}{n_0}&=\dfrac{R_{02}R_{10}+R_{02}R_{12}+R{12}R_{01}}{R_{10}R_{20}+R_{10}R_{21}+R_{12}R_{20}}\,,
    \end{split}
\end{equation}
where:
\begin{equation}\label{eq:three_level_rates}
    \begin{split}
        R_{10}&=n_ek_{10}+A_{10}/(1+0.5\tau_{10})\,,\\
        R_{20}&=n_ek_{20}+A_{20}\,,\\
        R_{21}&=n_ek_{21}+A_{21}/(1+0.5\tau_{20})\,,\\
        R_{01}&=(g_1/g_0)n_ek_{10}e^{-E_{10}/(kT)}\,,\\
        R_{02}&=(g_2/g_0)n_ek_{20}e^{-E_{20}/(kT)}\,,\\
        R_{12}&=(g_2/g_1)n_ek_{21}e^{-E_{21}/(kT)}\,.
    \end{split}
\end{equation}
Inside of each HII region we generally approximate $n_e = n_{\mathrm{H}} + n_{\mathrm{He}}$ throughout, assuming that helium is singly ionized along with hydrogen (although see \S \ref{sec:voiii_corr}). We ignore electrons from the ionization of heavier elements given their low abundances.\par
One needs to solve for the level populations self-consistently with the optical depths that enter Eq \ref{eq:three_level_rates}. The line center optical depth between levels $l$ and $u$ is given by: \citep{2011piim.book.....D}:
\begin{equation}\label{eq:taus}
\begin{split}
    \tau_{ul}&=\dfrac{g_u}{g_l}\dfrac{A_{ul}\lambda_{ul}^3}{4(2\pi)^{3/2}\sigma_V}n_lR_{\mathrm{out}}\left(1-\dfrac{n_ug_l}{n_lg_u}\right)\,,\\
    \sigma_V^2&=\dfrac{v_{\mathrm{th}}^2+v_{\mathrm{turb}}^2}{2}\,,\ v_{\mathrm{th}}=\sqrt{\dfrac{2kT}{m_{\mathrm{O}}}}\,.
\end{split}
\end{equation}
where the line width, $\sigma_V$, includes thermal and turbulent broadening components. We self-consistently find numerical solutions for the level populations and optical depths by searching through a grid in $\tau_{10}, \tau_{21}$.

After solving for the level populations (Eqs \ref{eq:level_ratios} and \ref{eq:three_level_rates}) and optical depths (Eq \ref{eq:taus}) the [\oiii] 88 $\mu$m emission is determined via a modified version of Eq \ref{eq:l10_lowb} as:
\begin{equation}\label{eq:l10_general}
\begin{split}
    L_{10}&=\dfrac{R_1}{1+R_1+R_2}\left(\dfrac{n_{\mathrm{O}}}{n_{\mathrm{H}}}\right)_\odot\dfrac{Z}{Z_\odot}\dfrac{A_{10}}{1+0.5\tau_{10}}h\nu_{10}\dfrac{Q_{\mathrm{HI}}}{\alpha_{\mathrm{B,HII}}n_e}\,.
\end{split}
\end{equation}
Combining Eq (\ref{eq:l10_general}) with the \cite{2003A&A...397..527S} Q-SFR model (Eq \ref{eq:q_sfr}) gives the following generalized $L_{10}$-SFR relation:
\begin{equation}\label{eq:l10sfr_general}
    \begin{split}
        \dfrac{L_{10}}{L_\odot}=&10^{10.86-0.0029(7.3+\log(Z/Z_\odot))^{2.5}}\dfrac{Z}{Z_\odot}\\
        &\times\dfrac{R_1}{1+R_1+R_2}\dfrac{1}{1+0.5\tau_{10}}\dfrac{\mathrm{[cm^{-3}]}}{n_e}\dfrac{\mathrm{SFR}}{M_\odot/\mathrm{yr}}\,,
    \end{split}
\end{equation}
where the numerical values here assume $T_4=10^4$ K and $(n_\mathrm{O}/n_{\mathrm{H}})_\odot=10^{-3.31}$. 

\subsubsection{Line broadening}\label{sec:broadening}

A potential concern for our model is that the radiative trapping effect, if important, will depend on the geometry and velocity distribution of the OIII ions. Neither of these features are well-described in our simplified model or the \textsc{CLOUDY} simulations. For example, discrete HII regions are less likely to ``shadow'' each other than implied by our spherically symmetric treatment. In addition the HII regions will be moving around in their host galaxy with some velocity dispersion, and so the line profile will not be set by pure thermal broadening. However, even in the pure thermal broadening case -- which likely overestimates radiative trapping effects -- we find that trapping is only important at relatively high $Q_{\mathrm{HI}}$, $n_{\mathrm{H}}$, and $Z$. A large value of $Q_{\mathrm{HI}}$ makes the Str$\ddot{\mathrm{o}}$mgren radius and optical depth large while the optical depth also increases with $n_{\mathrm{H}}$, $Z$. 

\begin{figure}
    \centering
    \includegraphics[width=0.5\textwidth]{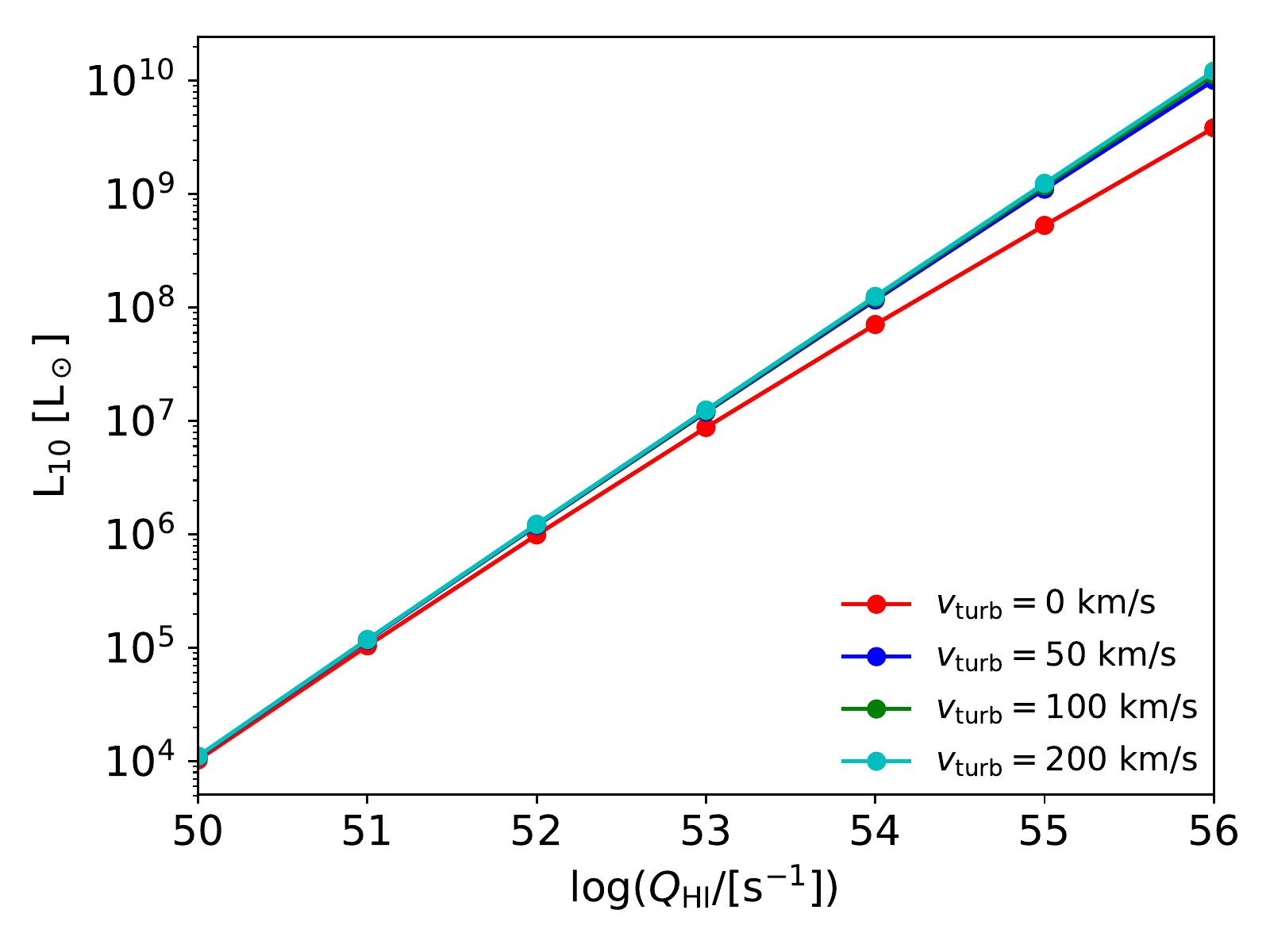}\\
    \caption{Radiative trapping and line broadening.
    The curves show \textsc{CLOUDY} results for $L_{\mathrm{10}}$ versus $Q_{\mathrm{HI}}$ assuming 
    $\log (n_{\mathrm{H}}/\mathrm{[cm^{-3}]})=2.0$, and $Z/Z_\odot=0.2$, for various choices of turbulent velocity broadening. In the pure thermal broadening case ($v_{\mathrm{turb}} = 0$ km/s), radiative trapping leads to a decline in $L_{10}$ at high $Q_{\mathrm{HI}}$.
    In the more likely case of additional line broadening, radiative trapping is a small effect and the results are insensitive to the precise choice of $v_{\mathrm{turb}}$.}\label{fig:v_turb}
\end{figure}

To roughly account for non-thermal motions, we simply add a turbulent velocity component to the velocity width in Eq \ref{eq:taus}. Our fiducial choice for $v_{\mathrm{turb}}$ is 100 km/s, although our results are insensitive to this choice as illustrated in Fig \ref{fig:v_turb}.\footnote{Treating the non-thermal motions as an additional velocity dispersion is imperfect because it ignores the fact that nearby OIII ions will have correlated velocities.} Radiative trapping is a small effect provided that additional broadening effects -- beyond pure thermal broadening -- impact the line profiles.

\subsubsection{Results with Eq \ref{eq:l10_general}}

\begin{figure*}
    \centering
    \includegraphics[width=1\textwidth]{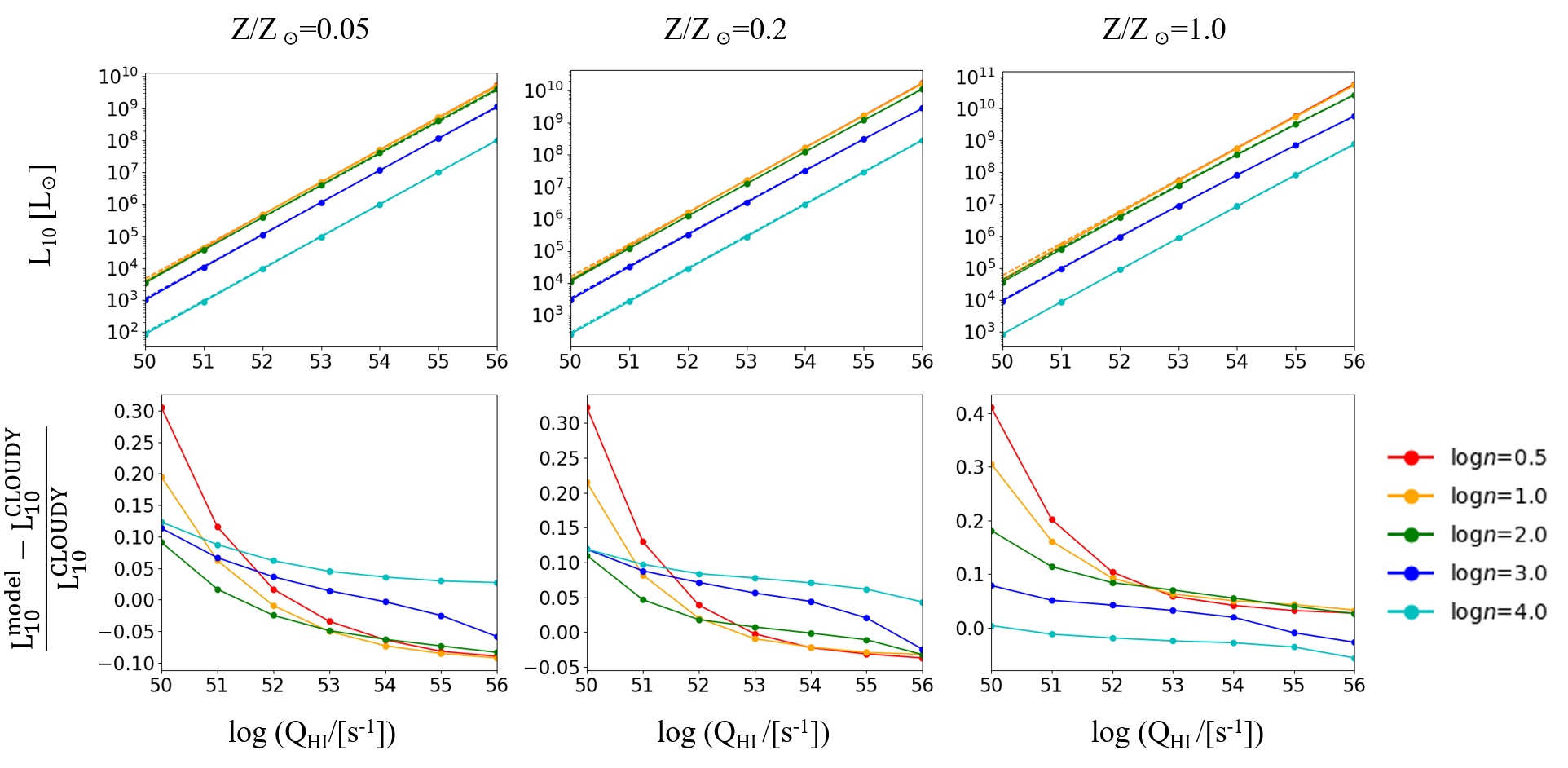}\\
    \caption{Comparison between $L_{10}$ predicted by Eq (\ref{eq:l10_general}) and \textsc{CLOUDY}. This figure is similar to Fig \ref{fig:l10low_sfr_accuracy} except here we account for a full three-level atomic treatment including collisional de-excitations and radiative trapping.
    }\label{fig:l10sfr_general}
\end{figure*}

Fig \ref{fig:l10sfr_general} compares the model of Eq \ref{eq:l10_general} with the corresponding \textsc{CLOUDY} simulations. 
This figure shows good overall accuracy, even at high densities $n_{\mathrm{H}} \gtrsim n_{\mathrm{crit}}$ where Eq \ref{eq:l10low_qhi} breaks down. Note that the model is again less accurate at high metallicity, with the $Z=Z_\odot$ showing larger differences with \textsc{CLOUDY}. This case is unlikely relevant for our high redshift galaxy science target. As in Fig \ref{fig:l10low_sfr_accuracy}, the model works less well at high metallicity owing to the increasing helium abundance with metallicity in our models.

\begin{figure}
    \centering
    \includegraphics[width=0.45\textwidth]{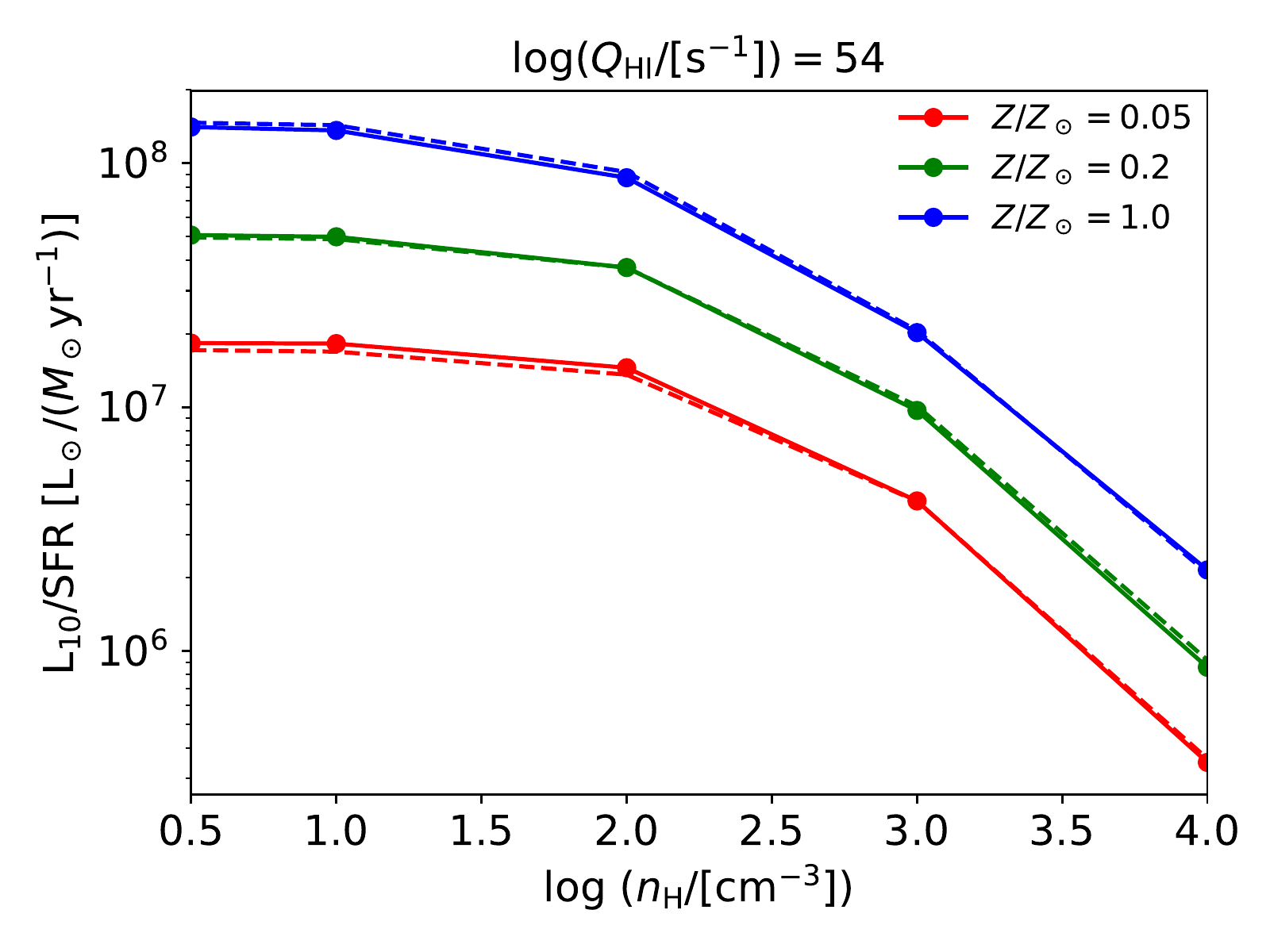}
    \caption{$L_{10}/\mathrm{SFR}$-$n_{\mathrm{H}}$ relation given by \textsc{CLOUDY} simulations (solid lines) and Eq \ref{eq:l10_general} (dashed lines) for different gas metallicities. We fix $Q_{\mathrm{HI}}=10^{54}$ s$^{-1}$ here, but the results are insensitive to the precise ionization rate adopted. The ratio drops with decreasing metallicity and increasing density, with the latter effect owing to collisional de-excitations.}\label{fig:LSFR_cloudy}
\end{figure}\par

Fig \ref{fig:LSFR_cloudy} shows explicitly how $L_{10}/\mathrm{SFR}$ declines with increasing $n_{\mathrm{H}}$ owing to collisional de-excitations. As expected, the drop-off starts around $n_e \sim 10^2-10^3$ cm$^{-3}$, comparable to the critical density ($n_{\mathrm{crit}} = 1.7 \times 10^3$ cm$^{-3}$ at $T_4=1$.) The good agreement with \textsc{CLOUDY} shows that this effect is well-captured by our three-level atomic treatment. The figure also illustrates how $L_{10}/\mathrm{SFR}$ drops as the gas metallicity decreases. Given the relatively large values of $L_{10}/\mathrm{SFR}$ from recent ALMA measurements, we will use these effects to bound the gas phase metallicity and HII region gas density at $z \sim 6-9$ in \S \ref{sec:ALMA_constraints}.

\subsection{$V_{\mathrm{OIII}}/V_{\mathrm{HII}}$ correction}\label{sec:voiii_corr}

In the previous sections, we assumed that the volumes of doubly-ionized oxygen and ionized hydrogen across a galaxy are identical with $V_{\mathrm{OIII}}/V_{\mathrm{HII}}=1$. In fact, this ratio is less than unity for galaxies with soft ionizing spectra and/or low rates of ionizing photon production.\footnote{Note that the volume of doubly ionized oxygen can never {\em exceed} that of ionized hydrogen. OII ionizing photons encountering neutral hydrogen gas just beyond the edge of an HII region have a much greater probability of being absorbed by abundant neutral hydrogen atoms than by OII and so these photons mostly go into growing the HII region.} In this section, we model the volume ratio, $V_{\mathrm{OIII}}/V_{\mathrm{HII}}$, and compare it with CLOUDY calculations. This allows us to ``correct'' our previous estimates and more accurately model OIII emission. 

One effect that can limit the size of doubly-ionized oxygen zones is absorption from neutral helium atoms. Absorption from neutral helium plays a small role in the case of fairly hard ionizing spectra, where the HeII and HII zones coincide. However, in the case of a soft spectrum an HII region will have an inner HeII zone, while the outer parts of the HII region will mostly be in the form of HeI.\footnote{Across the range of ionizing spectra considered in this work, it is a very good approximation to neglect doubly-ionized helium as well as higher ionization stages of oxygen (i.e., those beyond OIII).} In this soft spectrum case, the doubly-ionized oxygen zone is essentially set by the size of the HeII region. An estimate of the ratio of the volumes in the HeII and HII zones may be made by neglecting the absorption of HeI ionizing photons by HI and the absorption of HI ionizing photons by HeI. In this case, ignoring also the small increase in the free electron abundance in the HeII regions, the volume ratio is \citep{2006agna.book.....O}:

\begin{equation}\label{eq:heii_hii}
\frac{V_{\mathrm{HeII}}}{V_{\mathrm{HII}}} = \frac{Q_{\mathrm{HeI}}}{Q_{\mathrm{HI}}}\frac{n_{\mathrm{H}}}{n_\mathrm{He}}\frac{\alpha_\mathrm{B,HII}}{\alpha_{\mathrm{B,HeII}}}\,.
\end{equation}
Here $Q_{\mathrm{HeI}}$ denotes the HeI ionizing photon rate, i.e. the number of photons per second produced above the HeI photoionization threshold of $24.6$ eV, $n_\mathrm{He}$ is the number density of helium atoms, and $\alpha_\mathrm{B,HeII}$ is the HeII recombination coefficient. Eq \ref{eq:heii_hii} implies that the HeII and HII zones coincide with $V_{\mathrm{HeII}}/V_{\mathrm{HII}} = 1$ provided $Q_\mathrm{HeII}/Q_\mathrm{HI} \geq (n_\mathrm{He} \alpha_\mathrm{B,HeII})/(n_\mathrm{H} \alpha_\mathrm{B,HII})$. Assuming gas of primordial composition with a temperature of $T_4=1$, this requires $Q_\mathrm{HeI} \geq 0.086 Q_\mathrm{HI}$.
For our fiducial ionizing source spectra, this requirement is met and the HeII and HII zones match up. On the other hand, in the case of a softer source spectrum, we expect $V_\mathrm{OIII}/V_\mathrm{HII} \sim V_\mathrm{HeII}/V_\mathrm{HII}$. 

In the harder spectrum case, with coincident HeII/HII zones, we can model the OIII fraction as a function of radius across the HII region and compute the volume averaged OIII fraction. Even in this harder spectrum case, we find that the average OIII fraction may be significantly less than unity for certain plausible ionizing source and HII region parameters. In order to calculate the OIII fraction, we apply photoionization equilibrium accounting for OII photoionizations, OIII recombinations, and charge-exchange reactions between OIII ions and residual neutral hydrogen atoms within an HII region.
The latter process, in which an electron swaps from a hydrogen atom to an OIII ion to form OII (and an HII ion), can play an important role in the ionization balance in the outskirts of HII regions \citep{2006agna.book.....O}. The ionization balance equation may be written as:

\begin{equation}\label{eq:35}
    \dfrac{Q_{\mathrm{OII}}(0) \langle{e^{-\tau_\nu} \sigma_\mathrm{OII}(\nu)}\rangle}{4\pi R_{\mathrm{out}}^2y^2}n_{\mathrm{OII}}=n_{\mathrm{OIII}}n_e\alpha_{\mathrm{OIII}}+n_{\mathrm{OIII}}n_{\mathrm{HI}}\delta'.
\end{equation}
In this equation, $Q_\mathrm{OII}$ is the rate of OII ionizing photons produced by the galaxy:
\begin{equation}\label{eq:qoii}
    Q_{\mathrm{OII}}(y=0)=\int_{\nu_{\mathrm{OII}}}^\infty\dfrac{L_\nu}{h\nu}d\nu\,,
\end{equation}
here $\nu_{\mathrm{OII}}$ is the OII photoionization threshold at an energy of $35.12$ eV, $L_\nu$ is the spectral energy distribution of the central radiation source.
The left-hand side of Eq \ref{eq:35} describes the OII photoionization rate, with $y$ denoting the radius in units of the Str$\ddot{\mathrm{o}}$mgren radius, $R_\mathrm{out}$. The quantity $\langle{e^{-\tau_\nu} \sigma_\mathrm{OII}(\nu)\rangle}$ is a weighted average over the ionizing spectrum:
\begin{equation}\label{eq:weigthed_etauoii}
\langle{e^{-\tau_\nu} \sigma_\mathrm{OII}(\nu)\rangle} =
\frac{\int_{\nu_{\mathrm{OII}}}^\infty\dfrac{L_\nu}{h\nu} e^{-\tau_\nu} \sigma_\mathrm{OII}(\nu) d\nu}{\int_{\nu_{\mathrm{OII}}}^\infty\dfrac{L_\nu}{h\nu}d\nu}
\end{equation}
Here $\sigma_\mathrm{OII}$ is the OII photoionization cross section, and $e^{-\tau(\nu)}$ accounts for the absorption of OII ionizing photons. We calculate the optical depth $\tau_\nu(y)$ -- we generally suppress the $y$ dependence in our notation for brevity -- assuming that it is dominated by residual hydrogen absorption, neglecting contributions from helium and other elements. This is a good approximation, provided the criterion described around Eq \ref{eq:heii_hii} is met.

Finally, the terms on the right hand side of Eq \ref{eq:35} describe OIII recombinations and charge-exchange interactions, respectively. Here $\alpha_{\mathrm{OIII}}$ is the OIII recombination rate, and $\delta'=1.05\times 10^{-9}$ cm$^3$s$^{-1}$ is the charge exchange rate of the $\mathrm{OIII}+\mathrm{HI}\leftrightarrow \mathrm{OII}+\mathrm{HII}$ process near $T_4=1$ \citep{2006agna.book.....O}. The OIII recombination rate includes both radiative and dielectronic recombinations. Note that in order to account for absorption of OII ionizing photons by residual hydrogen, $e^{-\tau_\nu}$, and to compute the charge-exchange reaction rate we need to first compute the neutral hydrogen fraction $x_\mathrm{HI}(y)= n_\mathrm{HI}(y)/n_H$. This is accomplished by self-consistently solving for the photoionization balance of HI/HII and the optical depth, $\tau_\nu$. For simplicity, we again neglect the absorption of HI ionizing photons by helium and heavier elements. 
In this case, the optical depth equation is:
\begin{equation}\label{eq:28}
    \dfrac{d\tau_\nu}{dy}=n_{\mathrm{H}}R_{\mathrm{out}}\sigma_{\mathrm{HI}}(\nu) x_{\mathrm{HI}}\,,
\end{equation}
where $\sigma_{\mathrm{HI}}$ is the HI photoionization cross section. Photoionization equilibrium balance dictates that: 
\begin{equation}\label{eq:photo-balance}
    \dfrac{n_{\mathrm{HI}}}{4\pi r^2}\int_{\nu_\mathrm{HI}}^\infty\dfrac{L_\nu}{h\nu}\sigma_{\mathrm{HI}}(\nu)e^{-\tau_\nu}d\nu=n_{\mathrm{HII}}n_e\alpha_{\mathrm{B, HII}}(T).
\end{equation}
Eqs \ref{eq:28} and \ref{eq:photo-balance} can be solved numerically to determine $x_\mathrm{HI}(y)$ starting from the inner radius, $y=0.01$, at which point the HI optical depth is negligibly small. The resulting model for $x_\mathrm{HI}(y)$ agrees with more detailed CLOUDY calculations: for example, the volume-averaged ionized fractions match to within 3\%. The small differences owe to our neglect of helium absorption, and our approximate estimates of HII region temperatures which are also assumed to be independent of radius. 

After calculating the neutral hydrogen fraction as a function of radius, $y$, we can then solve the OII/OIII photoionization equilibrium equation (Eq \ref{eq:35}) for the OII and OIII abundances. We assume that oxygen is entirely in the form of OII and OIII and denote the fractional OIII and OII abundances by $x_\mathrm{OIII} = n_\mathrm{OIII}/n_\mathrm{O}$, $1 - x_\mathrm{OIII}$, respectively. Here we also ignore the contributions of helium (and heavier ions) to the free electron density, assuming $n_e=n_{\mathrm{H}}(1-x_{\mathrm{HI}})$. In this case, the solution to Eq \ref{eq:35} is:
\begin{equation}\label{eq:40}
    y^2\dfrac{x_{\mathrm{OIII}}}{1-x_{\mathrm{OIII}}}=\dfrac{Q_{\mathrm{OII}}(0)}{4\pi R_{\mathrm{out}}^2}\dfrac{\langle\sigma_{\mathrm{OII}}e^{-\tau_\nu}\rangle}{n_{\mathrm{H}}[(1-x_{\mathrm{HI}})\alpha_{\mathrm{OIII}}+x_{\mathrm{HI}}\delta']}\equiv A(y)\,.
\end{equation}
The resulting doubly-ionized oxygen fraction is therefore:
\begin{equation}\label{eq:xoiii}
    x_{\mathrm{OIII}}(y)=\dfrac{A(y)}{A(y)+y^2}\,.
\end{equation}\par
In the limit that $A(y) >> y^2$, $x_\mathrm{OIII} \rightarrow 1$ but $A(y)$ may fall below $y^2$ within the HII region for some parameters of interest. In the latter case, $x_\mathrm{OIII}$ will drop-off within the HII region.
It is also instructive to calculate $A(y)$ in the limit that $x_\mathrm{HI} \ll 1$, ignoring residual hydrogen absorption and charge-exchange reactions. In this limit, $A$ is independent of $y$ and can be written as:
\begin{equation}\label{eq:a_approx}
A \approx \frac{1}{3} \frac{Q_\mathrm{OII}(0)}{Q_\mathrm{HI}(0)} \frac{n_\mathrm{H}}{n_\mathrm{O}} \frac{\alpha_\mathrm{B,HII}}{\alpha_\mathrm{OIII}} n_\mathrm{O} \langle{\sigma_\mathrm{OII}\rangle} R_\mathrm{out}.
\end{equation}
One can see that (when $x_\mathrm{HI} \ll 1$) $A \propto (Q_\mathrm{OII}/Q_\mathrm{HI}) n_\mathrm{H}^{1/3} Q_\mathrm{HI}^{1/3}$. Note that the ionization parameter $U$ scales as $U \propto n_\mathrm{H}^{1/3} Q_\mathrm{HI}^{1/3}$, and so at small $y$ the quantity $A$ is set by the spectral shape and ionization parameter. Hence the volume correction becomes important when the first factor becomes small (soft spectrum), when the HII region has a low gas density, and/or a small ionizing luminosity. Furthermore, the absorption of OII ionizing photons by residual hydrogen and the charge-exchange reactions reduce $A$ as $y$ grows. On the other hand, in the case where the product of $Q_\mathrm{OII}/Q_\mathrm{HI}$, $n_\mathrm{H}$, and $Q_\mathrm{HI}$ is sufficiently large, $x_\mathrm{OIII} \sim 1$ is a good approximation across the entire region and $V_\mathrm{OIII} \sim V_\mathrm{HII}$.

\begin{figure}
    \centering
    \includegraphics[width=0.45\textwidth]{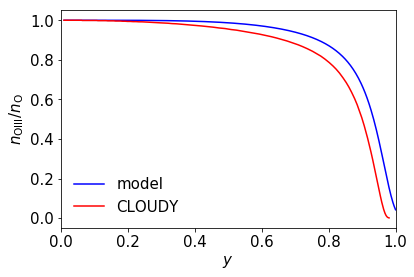}
    \caption{OIII radial abundance predicted by the numerical solution of Eq ({\ref{eq:40}}) and \textsc{CLOUDY} for an example case with $\log n_{\mathrm{H}}/[\mathrm{cm^{-3}}]=0.5$, $\log Q_{\mathrm{HI}}/\mathrm{s^{-1}}=50$ and $Z/Z_\odot=0.2$. The ionizing radiation spectrum between 1 and 4 Rydbergs is modeled as a blackbody with effective temperature $T_{4,\mathrm{eff}}=5.5$ K.}\label{fig:15}
\end{figure}\par

Fig \ref{fig:15} shows an example model calculation of $x_\mathrm{OIII}(y)$ in comparison to \textsc{CLOUDY}. In this example, we model the ionizing radiation spectrum between 1 and 4 Rydbergs as a blackbody with an effective temperature of $T_{\mathrm{4,eff}}=5.5$, which provides a good approximation to our fidutial model spectrum. The ISM parameters are $Q_\mathrm{HI}=10^{50}$ s$^{-1}$, $n_\mathrm{H} = 10^{0.5}$ cm$^{-3}$, and $Z=0.2 Z_\odot$. Note that the hydrogen photoionization rate here is much smaller than that expected for currently observed galaxies (\S \ref{sec:ALMA_constraints}): it is chosen to illustrate a case where the volume correction factor is relatively large. 
With these parameters, both the model and \textsc{CLOUDY} calculations show a steep drop starting near $y \gtrsim 0.8$. This is a consequence of the increasing residual neutral hydrogen abundance and charge-exchange reactions play an important role in the declining OIII abundance. One can see that our model captures the trend in \textsc{CLOUDY} well, with \textsc{CLOUDY} showing a slightly stronger decline. Tests with \textsc{CLOUDY} reveal that this mildly stronger trend owes largely to absorption from residual neutral helium within the HII region, which is neglected in our treatment. \par

Finally, we can use our computations of the doubly-ionized fraction throughout the HII region to determine the volume-averaged OIII fraction as:
\begin{equation}\label{eq:V_corr}
    \dfrac{V_{\mathrm{OIII}}}{V_{\mathrm{HII}}}=3\int_0^1x_{\mathrm{OIII}}(y)y^2dy\,.
\end{equation}
If the spectrum is so soft that the criterion described around Eq \ref{eq:heii_hii} is not satisfied, we instead use the approximation $V_\mathrm{OIII} \sim V_\mathrm{HeII}$. 
We further compare the accuracy of these calculations with a broad range of \textsc{CLOUDY} models. \par

\begin{figure}
    \centering
    \includegraphics[width=0.45\textwidth]{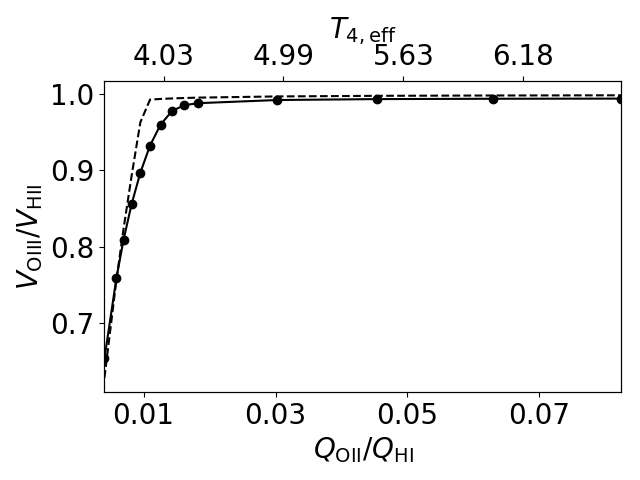}
    \includegraphics[width=0.45\textwidth]{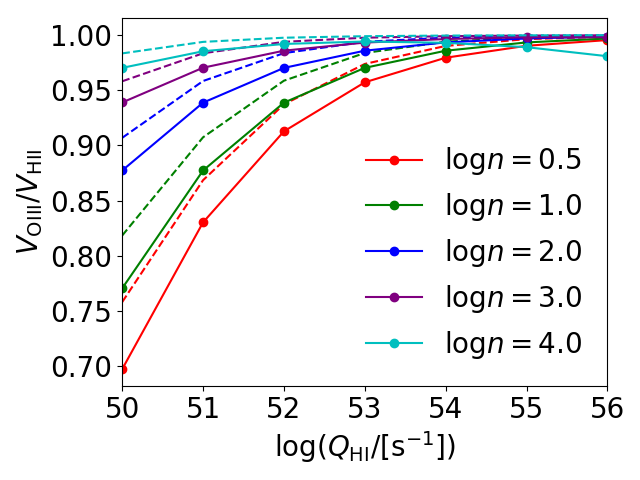}
    \caption{Variation of the OIII volume correction under changes in the spectrum of ionizing radiation, hydrogen density, and ionization rate. Solid lines show \textsc{CLOUDY} results, while the dashed lines show the OIII volume analytic model. {\em Top panel:} Changes with spectral shape, parameterized by an effective black-body temperature, $T_{4, \mathrm{eff}}.$ The other parameters are fixed at: $n_{\mathrm{H}}=100$ cm$^{-3}$, $Z=0.2Z_\odot$, $Q=10^{54}$ s$^{-1}$. {\em Bottom panel:} Variations with hydrogen density and the rate of hydrogen ionizing photon production. Here we fix $Z=0.2Z_\odot$ and $\mathrm{T_{4,eff}}=5.5$.}\label{fig:V_correction}
\end{figure}\par

The top panel of Fig \ref{fig:V_correction} shows how the volume correction factor varies with spectral shape. As motivated in \S \ref{sec:cloudy}, in order to treat variations around our fiducial \textsc{starburst99} spectrum we model the ionizing radiation spectrum between 1 and 4 Rydbergs
as a blackbody with an effective temperature of $T_{4,\mathrm{eff}}$. As mentioned earlier, $T_{\mathrm{4,eff}}=5.5$ matches our fiducial spectrum. 
Fig \ref{fig:V_correction} shows that the volume correction factor is near unity  for $Q_{\mathrm{HI}}=10^{54}$ s$^{-1}$ and $n_{\mathrm{H}}=100$ cm$^{-3}$ provided $T_{4,\mathrm{eff}} \gtrsim 4$. The volume correction factor ($V_{\mathrm{OIII}}/V_{\mathrm{HII}}$) falls sharply for softer ionizing spectra (i.e., for lower $T_{4,\mathrm{eff}}$) owing to helium, as anticipated in Eq \ref{eq:heii_hii}. Our estimate captures the location in temperature and the extent of this decline well. \par

In the context of population synthesis models, these results suggest that the volume correction (for sufficiently high $Q_{\mathrm{HI}}$, $n_{\mathrm{H}}$ as explored in the next paragraph) is a small effect for galaxies with ongoing or very recent star formation. More specifically, we consider continous SFR \textsc{starburst99} models of varying stellar age and metallicity. The spectrum softens with increasing metallicity and age, before reaching an asymptotic shape around 10 Myr. However, in these cases the spectrum is still harder than the $T_{4,\mathrm{eff}} \lesssim 4$ case where the volume correction becomes important. Quantitatively, we find $Q_{\mathrm{HeI}}/Q_{\mathrm{HI}} \gtrsim 0.20$ for $Z=0.2 Z_\odot$ for continous SFR models. Furthermore, for a stellar age of 10 Myr, $Q_{\mathrm{HeI}}/Q_{\mathrm{HI}} \gtrsim 0.15$ for $Z \leq Z_\odot$. That is, the criterion of Eq \ref{eq:heii_hii} is satisfied for plausible continuous SFR models. The volume correction would, however, be more important for galaxies with aging starbursts and age $\gtrsim 5$ Myr (at which point $T_{4,\mathrm{eff}} \leq 4$ for $Z=0.2 Z_\odot$ in the \textsc{starburst99} models), although binarity and rapid rotation may naturally produce harder spectra even for aging starburst models \citep{Stanway14}.


In the bottom panel of Fig \ref{fig:V_correction} we show how the volume correction factor depends on $Q_{\mathrm{HI}}$ and $n_{\mathrm{H}}$. As discussed earlier, the volume correction factor becomes more important at low ionization parameter or equivalently in the low $Q_{\mathrm{HI}}$, $n_{\mathrm{H}}$ part of the figure. The trend seen in the \textsc{CLOUDY} calculations is well matched in our model, although the volume corrections in the simulations are slightly more important than predicted.\footnote{The \textsc{CLOUDY} simulations show a slight turndown at high $Q_{\mathrm{HI}}$ and $n_{\mathrm{H}}$ that is not captured in our model. This likely owes to an additional charge-exchange process, $\mathrm{OIII+HII\leftrightarrow OIV+HI}$, which is not included in our model but occurs in the inner part of the OIII region at high $Q_{\mathrm{HI}}$, $n_{\mathrm{H}}$.}   
Note that while the low $Q_{\mathrm{HI}}$ regime -- where the correction becomes more important -- 
is irrelevant for interpreting the current ALMA observations, it may be of interest for future line-intensity mapping applications. 

\begin{figure*}
    \centering
    \includegraphics[width=1.0\textwidth]{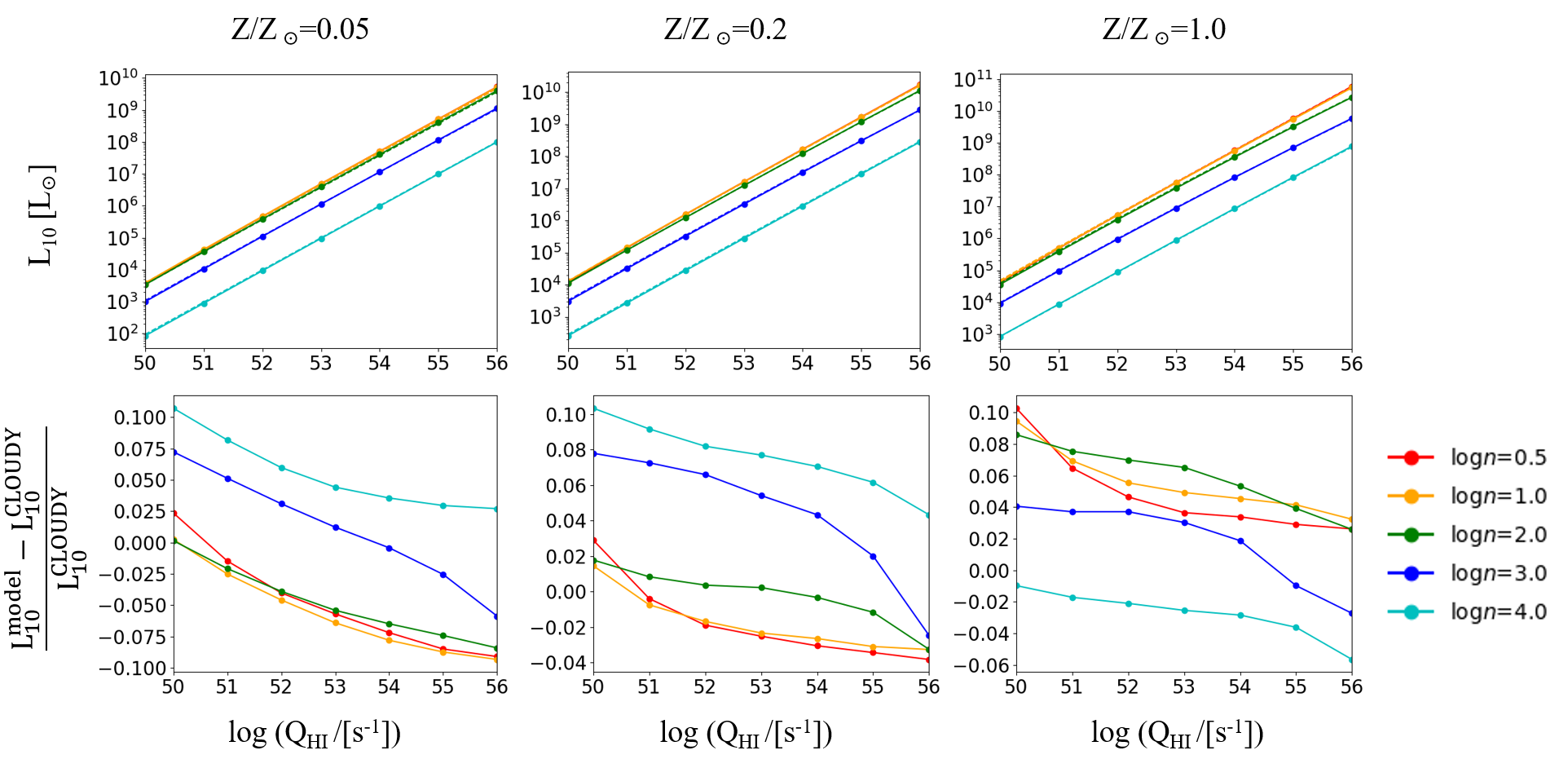}
    \caption{$L_{10}$ predicted through combining the [\oiii] model introduced in Eq (\ref{eq:l10_general}) and the OIII volume correction factor introduced in Eq (\ref{eq:V_corr}) compared with \textsc{CLOUDY} simulation results.
    Otherwise this figure is similar to Figs \ref{fig:l10sfr_general} and \ref{fig:l10low_sfr_accuracy}.}\label{fig:l10sfr_general_vcorr}
\end{figure*}

Fig \ref{fig:l10sfr_general_vcorr} summarizes our comparison with the \textsc{CLOUDY} simulations. Here we extend our previous results (Fig \ref{fig:l10sfr_general}) to 
include the volume correction into our model. For our fiducial spectral shape, the volume correction improves the accuracy at low $Q_{\mathrm{HI}}$ with $\sim 10\%$ accuracy obtained over much of the parameter space studied.

\subsection{Probability Distributions}\label{sec:pdfs}

Up to this point, our model has assumed uniform HII regions and a uniform ionizing radiation field. This neglects scatter in the HII region properties and radiation field across each galaxy, and -- when applied to an ensemble of galaxies -- it ignores variations from galaxy to galaxy. Furthermore, it does not account for any radial dependence of the gas density, metallicity, etc. across individual HII regions. We can improve on these shortcomings by allowing some of the key properties to be drawn from probability distribution functions (PDFs). Specifically, we will allow for variations in the gas density ($n_{\mathrm{H}}$) and the metallicity ($Z$). We still adopt a single temperature and spectral shape in order to avoid adding further complexity to our model and since these quantities play a less important role in determining the [\oiii] 88 $\mu$m luminosity ($L_{10}$). In what follows, these are generally computed assuming the stellar metallicity matches the gas-phase metallicity.

It is instructive to first consider the low density limit $n_{\mathrm{H}} \ll n_{\mathrm{crit}}$, as in \S \ref{sec:oiii_lownh} except now allowing for variations in $n_{\mathrm{H}}$ and $Z$. It is straightforward to show that
in this limit, Eq \ref{eq:l10low_qhi} is modified as:
\begin{equation}\label{eq:l10_lownh_pdf}
    \langle{L_{10}\rangle}=\left(\dfrac{n_{\mathrm{O}}}{n_{\mathrm{H}}}\right)_\odot h\nu_{10}(k_{01}+k_{02})\dfrac{Q_{\mathrm{HI}}}{\alpha_{\mathrm{B,HII}}}\dfrac{\langle Z/Z_\odot n_{\mathrm{H}}n_e\rangle}{\langle n_{\mathrm{H}}n_e\rangle}\,,
\end{equation}
where the averages here denote ensemble averages over the HII regions in question. This equation is general and the averages could apply over a single HII region (with well behaved radial profiles in $n_{\mathrm{H}}$ and/or $Z$), over an ensemble of HII regions across a single galaxy, or across a set of different galaxies. This equation has a simple interpretation. Allowing for variations, the metallicity of Eq \ref{eq:l10low_qhi} is simply generalized to an $n_{\mathrm{H}}^2$-weighted average metallicity. In the case that the metallicity and $n_{\mathrm{H}}$ variations are uncorrelated, then $\langle{Z/Z_\odot n_{\mathrm{H}}^2\rangle}/\langle{n_{\mathrm{H}}^2\rangle} \rightarrow \langle{Z/Z_\odot\rangle}$, and our earlier equation (Eq \ref{eq:l10low_qhi}) holds with the metallicity replaced by the ensemble-averaged value. Note that in this low density case, Eq \ref{eq:l10_lownh_pdf} follows without specifying a particular form for the density/metallicity PDFs.  

The situation is more complicated in the arbitrary density case. Here we allow $n_{\mathrm{H}}$ and $Z$ to vary from HII region to HII region across a galaxy, adopting lognormal distributions for $n_{\mathrm{H}}$ and $Z$ to roughly account for variations in these quantities. In the arbitary density case, we assume that $n_{\mathrm{H}}$ and $Z$ vary independently (i.e., are uncorrelated) for simplicity. 

Specifically, we assume PDFs of the following forms:
\begin{equation}\label{eq:pdfs}
    \begin{split}
        \dfrac{dP}{d\log n_{\mathrm{H}}}&=\dfrac{1}{\sigma_1\sqrt{2\pi}}\exp\left[-\dfrac{1}{2}\left(\dfrac{\log n_{\mathrm{H}}-\mu_1}{\sigma_1}\right)^2\right]\,,\\
        \dfrac{dP}{d\log Z}&=\dfrac{1}{\sigma_2\sqrt{2\pi}}\exp\left[-\dfrac{1}{2}\left(\dfrac{\log n_{\mathrm{H}}-\mu_2}{\sigma_2}\right)^2\right]\,,
    \end{split}
\end{equation}
where $\{\mu_1,\sigma_1\}$ and $\{\mu_2,\sigma_2\}$ describe the mean and standard deviation of $\log(n_{\mathrm{H}}/[\mathrm{cm^{-3}}])$ and $\log Z/Z_\odot$, respectively. The mean of $n_{\mathrm{H}}$ and $Z/Z_\odot$ are determined, respectively, by: $\log\langle n_{\mathrm{H}}\rangle=\mu_1+ \mathrm{ln(10)} \dfrac{\sigma_1^2}{2}$ and $\log\langle Z\rangle=\mu_2+ \mathrm{ln(10)} \dfrac{\sigma_2^2}{2}$. We can then determine the ensemble-averaged luminosity, essentially from Eq \ref{eq:l10_general} and \S \ref{sec:voiii_corr}. One subtlety here, however, is that the escape probability and volume correction are already galaxy-averaged quantities. Therefore, we fix the escape probability and volume correction at their mean-density values -- effectively assuming a homogeneous medium in computing these two quantities, which in any case have only a minor impact on the luminosity -- while integrating over the lognormal PDFs to model local variations in the gas density and metallicity:
\begin{equation}\label{eq:l10_avg}
\begin{split}
    \langle{L_{10}\rangle}&=\dfrac{h\nu_{10}A_{10}\langle Z\rangle_{\log Z}}{1+0.5\tau_{10}}\dfrac{Q_{\mathrm{HI}}}{\alpha_{\mathrm{B,HII}}}\dfrac{\langle\frac{R_1n_{\mathrm{H}}}{1+R_1+R_2}\rangle_{\log n_{\mathrm{H}}}}{\langle n_{\mathrm{H}}n_e\rangle_{\log n_{\mathrm{H}}}}\,.
\end{split}
\end{equation}
Here $\langle\rangle_{\log n_{\mathrm{H}}}$ and $\langle\rangle_{\log Z}$ denote averages over the $\log n_{\mathrm{H}}$ and $\log Z$ distributions. The values of $\tau_{10}$ and $\tau_{20}$ are computed using the mean gas density and metallicity (as are the temperature and spectral shape). In this calculation, we further approximate $n_e$ by $n_e=1.08 n_H$ assuming that helium is mostly singly-ionized in the HII region and of primordial composition. Under this approximation, $L_{10}$ is linear in $Z$ and so the average luminosity is given by that at the mean metallicity. It is therefore independent of $\sigma_2$. On the other hand, the luminosity is a non-linear function of $n_{\mathrm{H}}$ and so the average luminosity generally differs from that at the mean gas density. We compare this ensemble-averaged expression with observational data.

\section{ALMA observational constraints}\label{sec:ALMA_constraints}

Here we derive constraints on the metallicity and gas density from the current sample of nine $z \sim 6-9$ LBGs with [\oiii] 88 $\mu$m luminosity measurements from \cite{2016Sci...352.1559I,2017A&A...605A..42C,2017ApJ...837L..21L,2018Natur.557..392H,2019PASJ...71...71H,2019ApJ...874...27T,2019arXiv191010927H}.
We use the compilation of data in \cite{Harikane19}. This sample consists of nine LBGs, photometrically identified with the Hubble Space Telescope or the Hyper-Suprime Camera on the Subaru telescope, and subsequently detected in the [\oiii] 88 $\mu$m line with ALMA. The SFRs are derived from measurements of the UV and IR luminosities.
The galaxies in the sample span a range of properties, with the SFRs varying from roughly $\sim 5-200 M_\odot/\mathrm{yr}$.
We compare the reported measurements of $L_{10}/\mathrm{SFR}$ with a small extension to Eq \ref{eq:l10_avg} to connect the ionization rate and SFR, employing the $Q_{\mathrm{HI}}-\mathrm{SFR}$ relation of Eq \ref{eq:q_sfr}. We start by assuming that the stellar metallicity is identical to the gas metallicity.  

Before deriving confidence intervals on the model parameters, it is instructive to compare the measurements from Table 1 and Table 2 of \cite{2019arXiv191010927H} with example models. 
Note that the model $L_{10}$ is nearly a linear function in the SFR and so $L_{10}/\mathrm{SFR}$ is close to constant. As discussed previously, the value of $L_{10}/\mathrm{SFR}$ declines with decreasing metallicity and when the gas density becomes larger than the critical density. The ALMA data show considerable scatter in $L_{10}/\mathrm{SFR}$ beyond that from measurement errors: this intrinsic scatter is likely the consequence of galaxy-to-galaxy variations in the gas metallicity and density. (This is above and beyond the variations from HII to HII region captured in Eq \ref{eq:l10_avg}.) 
\begin{figure*}
    \centering
    \includegraphics[width=0.49\textwidth]{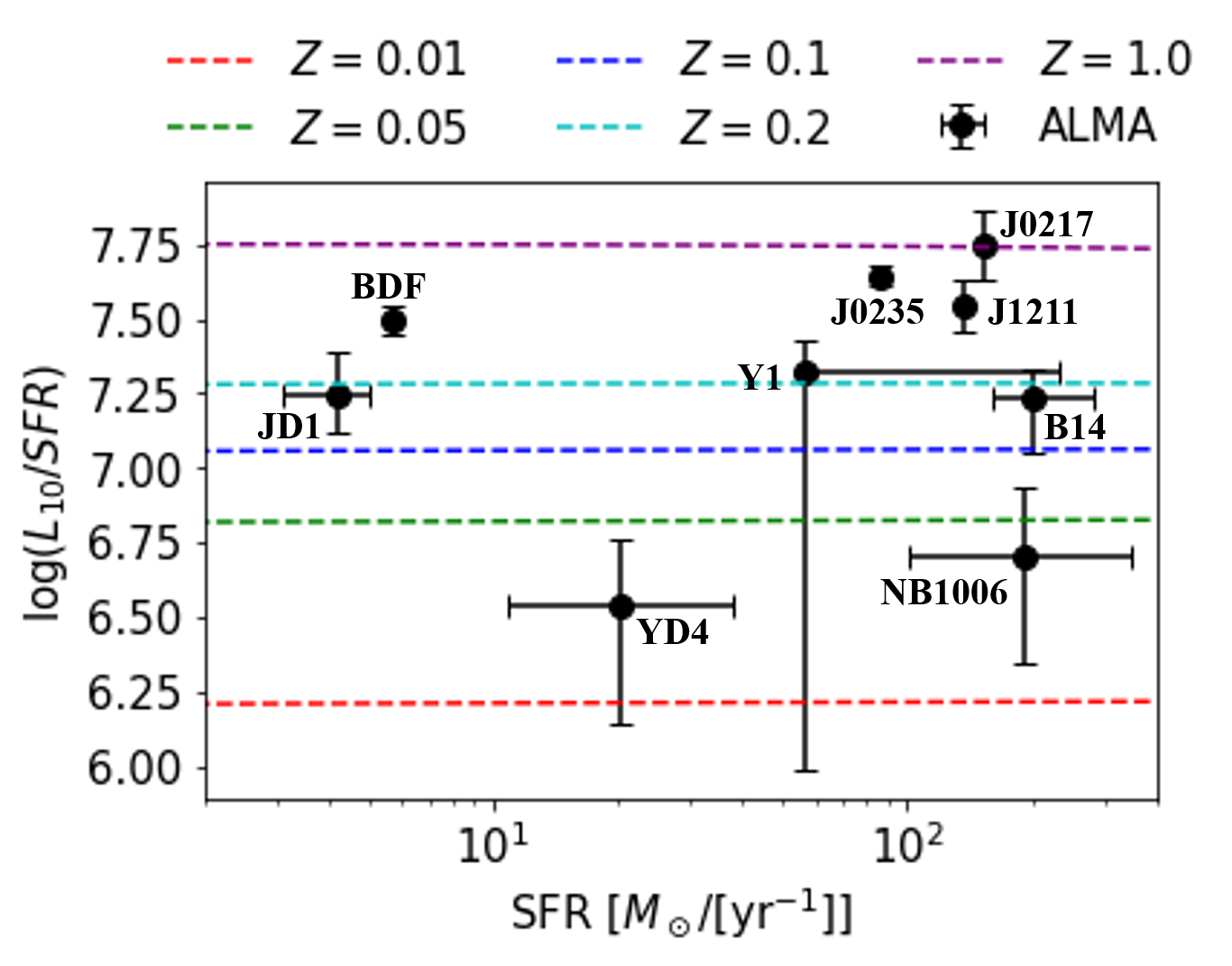}
    \includegraphics[width=0.49\textwidth]{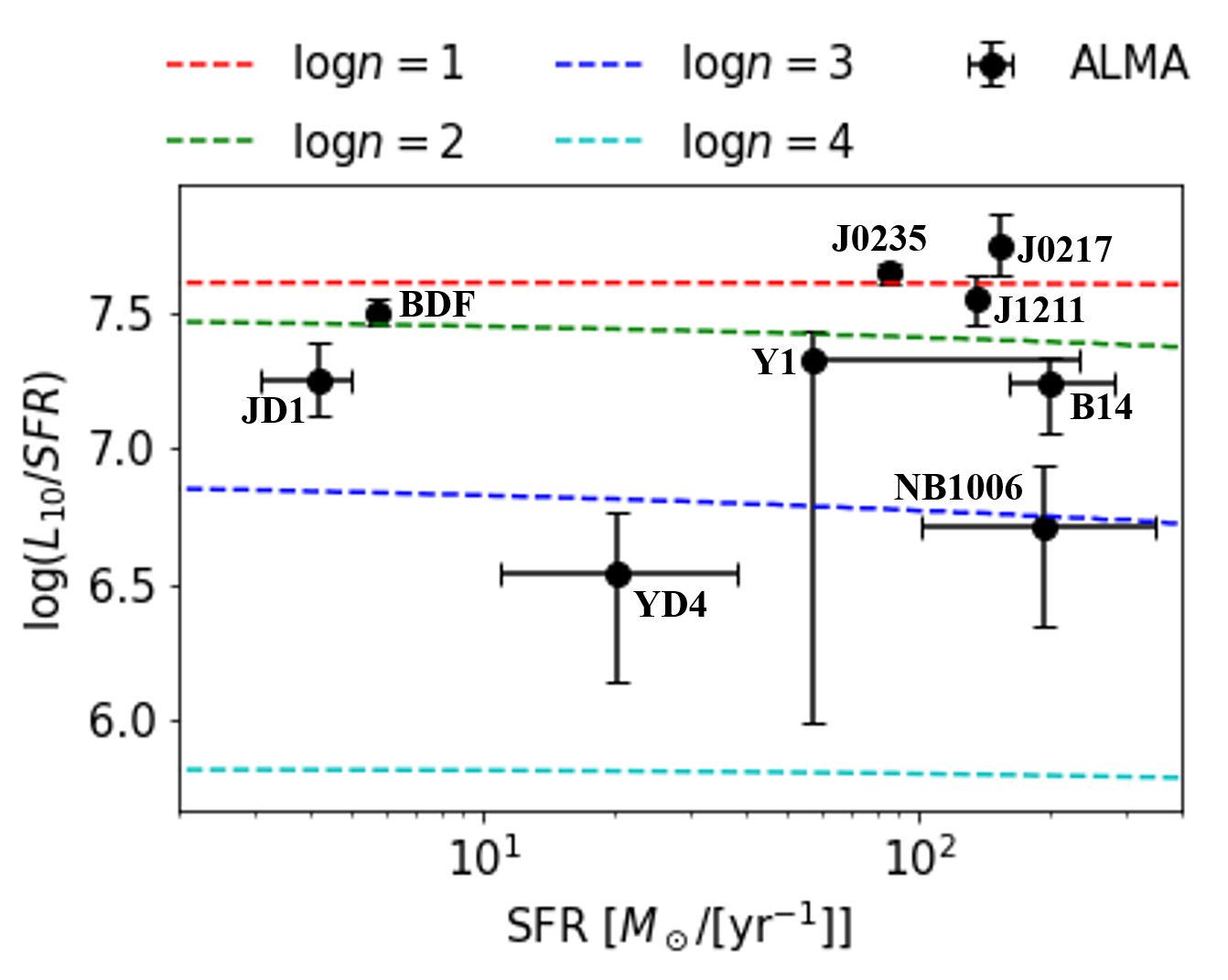}
    \caption{Comparison between the $L_{10}/\mathrm{SFR}$ versus SFR measurements from \protect\cite{Harikane19} and
    example models. The black points show the measurements with $1-\sigma$ error bars, while the dashed lines give model predictions according to Eq \ref{eq:l10_general}. Here
     $\log n$ in the legend specifies hydrogen density $\log(n_{\mathrm{H}}/[\mathrm{cm^{-3}}])$, while the metallicities, $Z$, are in solar units $Z/Z_\odot$. 
     {\em Left panel:} Models with varying metallicity; the hydrogen density is fixed to a low density limiting case value of $n_{\mathrm{H}}=10 \mathrm{cm^{-3}}$. {\em Right panel:} Models with varying hydrogen density; the metallicity is fixed to  $Z=0.6Z_\odot$.}\label{fig:datavsmodel}
\end{figure*}

Interestingly, as illustrated in Fig \ref{fig:datavsmodel}, there are four galaxies with $L_{10}/\mathrm{SFR} \gtrsim 10^{7.50} L_\odot/M_\odot/\mathrm{yr}$. Comparing to the example models suggests that these galaxies are best explained by metallicities of $Z=0.3-1.0 Z_\odot$ and relatively low gas densities, $n_{\mathrm{H}} \lesssim 10^{2}$ cm$^{-3}$. 

On the other hand, the two galaxies with the smallest $L_{10}/\mathrm{SFR}$ from the current sample -- centered around $L_{10}/\mathrm{SFR} \sim 10^{6.50} L_\odot/M_\odot/\mathrm{yr}$ -- are best explained by lower metallicities and/or higher gas densities. For example, the preferred metallicity for each of these two galaxies is around $Z \sim 0.03 Z_\odot$ in the low density limit (e.g. Fig \ref{fig:datavsmodel} shows $n_{\mathrm{H}} \sim 10$ cm$^{-3}$) or the gas density might be larger, $n_{\mathrm{H}} \sim 10^{3.2}$ cm$^{-3}$, if $Z=0.6 Z_\odot$.

To make these statements more precise, we need to translate the measured range in $L_{10}/\mathrm{SFR}$ into confidence intervals in metallicity and gas density. Given the large observed spread in $L_{10}/\mathrm{SFR}$, we determine separate bounds for each galaxy, rather than fitting to the ensemble of galaxies. One technical issue in deriving these bounds is that the $L_{10}/\mathrm{SFR}$ measurement errors for some of the galaxies are highly asymmetric; this appears to be a consequence of uncertainties in spectral energy distribution modeling \citep{2019ApJ...874...27T}.
To estimate confidence intervals, we need to adopt a plausible non-Gaussian model for the PDF of $L_{10}/\mathrm{SFR}$ that allows for strong asymmetry. We assume that the PDF for each galaxy follows an 
 epsilon-skew-normal distribution (ESN) distribution \citep{ESN}:
\[
  f(x)= 
\begin{cases}
    \dfrac{1}{\sqrt{2\pi}\sigma}\exp \left(-\dfrac{(x-\theta)^2}{2(1+\epsilon)^2\sigma^2}\right),&\text{if } x<\theta\\
    \dfrac{1}{\sqrt{2\pi}\sigma}\exp \left(-\dfrac{(x-\theta)^2}{2(1-\epsilon)^2\sigma^2}\right),& \text{if } x\geq\theta
\end{cases}
\]
where $x=\langle{L_{10}\rangle}/\mathrm{SFR}$ is given by Eq (\ref{eq:l10_avg}). To specify the parameters of this PDF, we make use of the ESN quantile function, which is:
\[
  q(u)= 
\begin{cases}
    \theta+\sigma(1+\epsilon)\Phi^{-1}\left(\dfrac{u}{1+\epsilon}\right),&\text{if } 0<u<(1+\epsilon)/2\\
    \theta+\sigma(1-\epsilon)\Phi^{-1}\left(\dfrac{u-\epsilon}{1-\epsilon}\right),& \text{if } (1+\epsilon)/2\leq u<1.
\end{cases}
\]
Here $u$ is the probability that the measured value of $L_{10}/\mathrm{SFR}$ is smaller than $q$ and $\Phi^{-1}$ is the inverse of the cumulative distribution function for the case of a Gaussian PDF. We solve for the ESN parameters $\theta$, $\sigma$, $\epsilon$ for each ALMA [\oiii]-emitting galaxy using:
\begin{equation}
    \begin{split}
        q(0.16)&=\hat{x}-\sigma^-\,,\\
        \theta&=\hat{x}\,,\\
        q(0.84)&=\hat{x}+\sigma^+\,,
    \end{split}
\end{equation}
where $\hat{x}$ is the $L_{10}/\mathrm{SFR}$ measurement, and $\sigma^{+/-}$ are the reported upper/lower measurement errors on $L_{10}/\mathrm{SFR}$ at 68\% confidence level.

\begin{table*}
\centering
\begin{tabular}{lcccc}
\hline
              & \multicolumn{2}{c}{$\sigma_1=0$}                  & \multicolumn{2}{c}{$\sigma_1=0.5$}                      \\
              & $\log(Z/Z_\odot)$ lower bound                  & $\log(n_{\mathrm{H}}/\mathrm{[cm^{-3}]})$ upper bound            & $\log\langle Z/Z_\odot\rangle$ lower bound                      & $\log\langle n_{\mathrm{H}}/\mathrm{[cm^{-3}]}\rangle$ upper bound               \\
&&&& \\[-1em]
\hline
&&&&\\[-1em]
J0217    &-0.23,-0.38&1.4,1.8&-0.18,-0.32&0.72,1.0\\
&&&&\\[-1em]
J0235 &-0.21,-0.27&1.5,1.7&-0.12,-0.19&0.68,0.86  \\
&&&&\\[-1em]
J1211 &-0.41,-0.54&1.8,2.1&-0.32,-0.46&1.0,1.2   \\
&&&&\\[-1em]
BDF &-0.42,-0.49&2.0,2.2&-0.33,-0.42&1.2,1.4\\
&&&&\\[-1em]
Y1 &-1.9,-2.8&3.0,3.8&-1.8,-2.8&2.5,3.5\\
&&&&\\[-1em]
JD1 &-0.81,-1.0&2.4,2.8&-0.73,-0.94&1.6,2.0  \\
&&&&\\[-1em]
B14 &-0.90,-1.1&2.3,2.7&-0.82,-1.0&1.6,1.9 \\
&&&&\\[-1em]
NB1006 &-1.6,-2.0&3.0,3.5&-1.6,-2.0&2.3,2.9 \\
&&&&\\[-1em]
YD4 &-1.8,-2.2&3.2,3.7&-1.7,-2.2&2.6,3.1\\
&&&&\\[-1em]
\hline
\end{tabular}
\caption{Summary of the $1-\sigma$ and $2-\sigma$ lower bounds on gas metallicity and upper bounds on gas density from [\oiii] emitting galaxies. The first (second) entry shows the $1-\sigma$ ($2-\sigma$) bound on the ISM parameters. The left-hand columns give results ignoring scatter from HII region to HII region across each galaxy, while the right-hand columns incorporate $\sigma_1=0.5$ dex of scatter.}\label{tb:2}
\end{table*}

To start with, we ignore the scatter from HII region to HII region in $n_{\mathrm{H}}$ and $Z$. Note that even in this simplified case we describe a single data point ($L_{10}/\mathrm{SFR}$ for each galaxy) with a two parameter model (fit separately for each galaxy). In a purely linear model this is an undetermined system. Here, we can nevertheless provide sensible bounds by adopting reasonable priors on $n_{\mathrm H}$ and $Z$. We adopt flat priors with $0\leq\log (n_{\mathrm{H}}/\mathrm{[cm^{-3}]})\leq4.0$ and $-5\leq\log (Z/Z_\odot)\leq0$. Furthermore, note that the model is not strictly linear in the model parameters: in particular, it saturates at low density, $n_{\mathrm{H}}$.\footnote{In principle, $L_{10}$ also saturates at high metallicity owing to radiative trapping (since $L_{10} \propto Z/\tau_{10}$ at high optical depth with $\tau_{10} \propto Z$), but this limit is not reached under our prior that $Z \leq Z_\odot$ for $v_{\mathrm{turb}}=100$ km/s.}

\begin{figure*}
    \centering
    \includegraphics[width=0.49\textwidth]{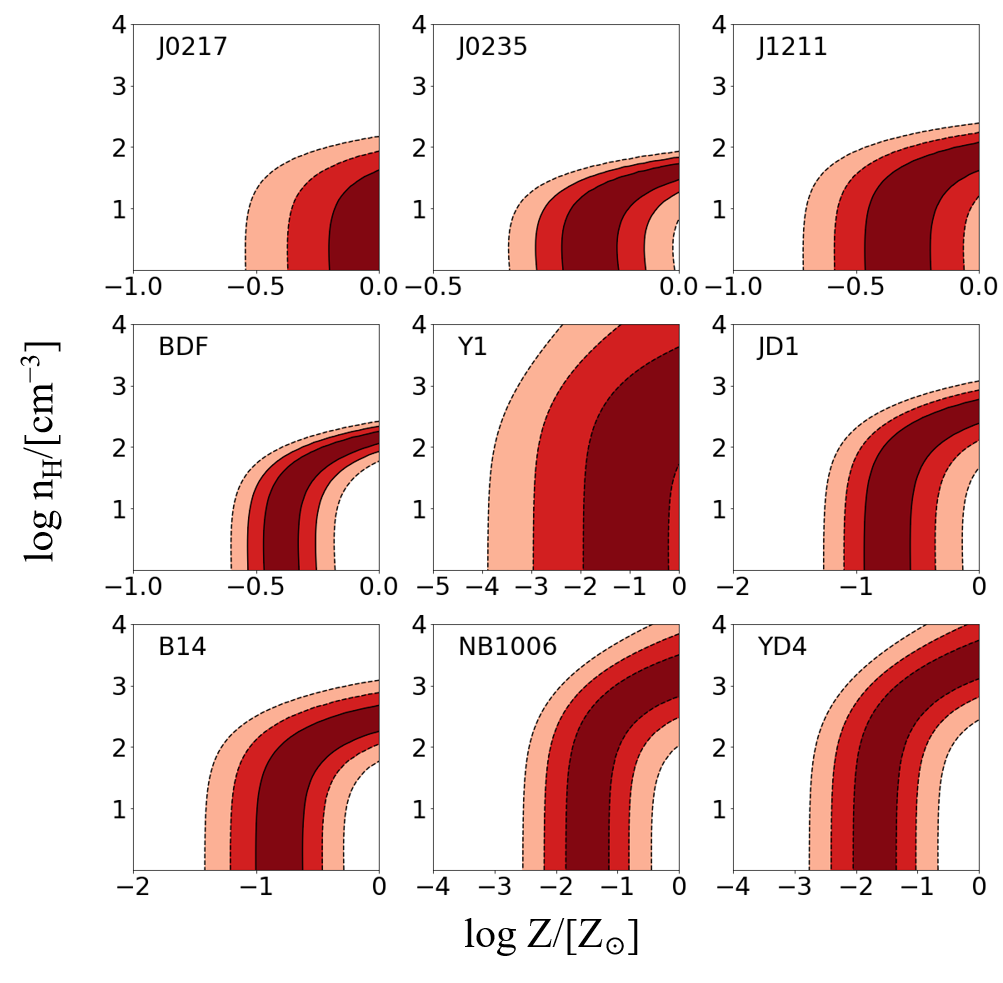}
    \includegraphics[width=0.49\textwidth]{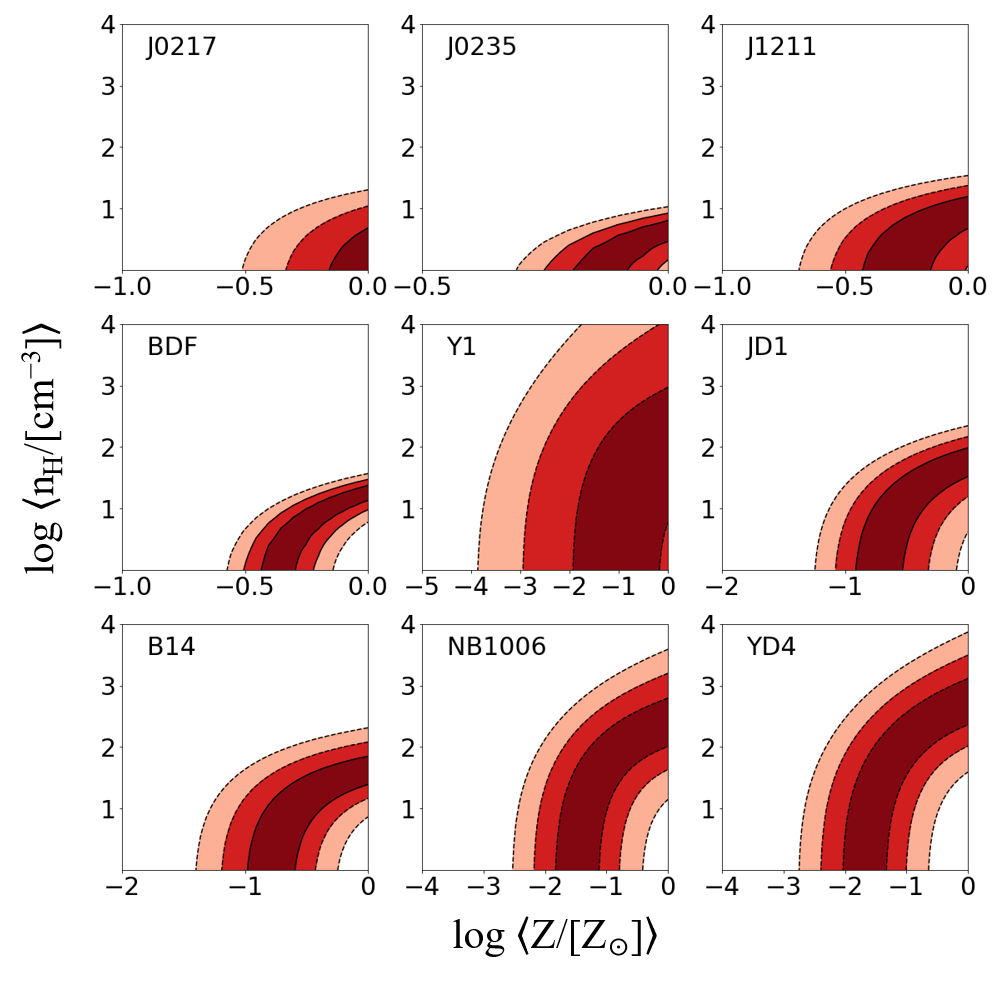}
    \caption{Gas metallicity and density constraints from the current ALMA measurements. {\em Left panel:} Here the results assume that the metallicity and gas density are uniform across each galaxy. The dark red, light red, and tan regions respectively enclose 68\%, 95\%, and 99.7\% confidence intervals.
    {\em Right panel:} Mean metallicity and density constraints allowing 0.5 dex of lognormal scatter in these quantities across each galaxy, as modeled using Eq \ref{eq:l10_avg}.
    }\label{fig:2Dposterior}
\end{figure*}

In the {\em left panel} of Fig \ref{fig:2Dposterior} we show the resulting 2D posteriors, in the $\log n_{\mathrm{H}}-\log Z$ plane,  for each [\oiii] emitting galaxy. The results illustrate the expected degeneracy between gas density and metallicity, with the allowed region in $n_{\mathrm{H}}$ and $Z$ for each galaxy resembling an ``L''-shape on its side. The measured values of $L_{10}/\mathrm{SFR}$ can be matched at either lower metallicity and gas density or higher metallicity and gas density. Again, this results because the suppression in luminosity from collisional de-excitations may be compensated by enhanced metallicities.  
However, the measurements yield interesting lower bounds on metallicity and upper bounds on the gas density. As noted earlier, these bounds vary considerably from galaxy to galaxy. 

Table \ref{tb:2} summarizes the 1 and 2-$\sigma$ bounds on metallicity and gas density. As discussed more qualitatively earlier, some of the galaxies in the sample are quite metal rich. For example, the $1-\sigma$ ($2-\sigma$) lower bound on the metallicity for J0217, at redshift 6.2, is $0.59 Z_\odot$ ($0.42 Z_\odot$). On the other hand, the lower bound on metallicity is rather loose for some of the galaxies in the sample, such as Y1 at redshift 8.3, for which $Z \geq 1.2 \times 10^{-2} Z_\odot$ and $Z \geq 1.6 \times 10^{-3} Z_\odot$ at $1-\sigma$ and $2-\sigma$ confidence, respectively. Likewise, the gas density bounds are variable with $n_{\mathrm{H}} \leq 25 \, (63)$ cm$^{-3}$ at $1-\sigma$ ($2-\sigma$) confidence for J0217, but $n_{\mathrm{H}} \leq 1.0 \times 10^3 (6.3 \times 10^3)$ cm$^{-3}$ at $1-\sigma$ ($2-\sigma$) for Y1. Note also that if we are agnostic to within our stated priors on metallicity and gas density, all of the galaxies in the sample are consistent with solar metallicity, for example, as well as low gas densities. This is illustrated explicitly in Fig \ref{fig:2Dposterior} and arises because -- as emphasized previously -- the increase in luminosity at high metallicity may be counteracted by collisional de-excitations at larger densities. Similarly, if the metallicity is relatively small then the data can accommodate arbitrarily low densities where collisional de-excitations are unimportant. We believe the most robust constraints from current data are our lower bounds on metallicity and upper bounds on gas density.
In any case, the ALMA measurements are clearly starting to provide interesting constraints on ISM properties. 

In order to explore how sensitive these results are to the assumption of uniform metallicity and density, we allow scatter in these quantities as described by Eq \ref{eq:pdfs}-\ref{eq:l10_avg}. As mentioned earlier, since $L_{10}$ is roughly a linear function of metallicity, these results do not depend on the width assumed for the metallicity distribution. Given that the present data sample is relatively small, we simply fix the width of the hydrogen density distribution at $\sigma_1=0.5$ dex. Although this value is arbitrary, it nevertheless gives a sense for the impact of plausible density variations. The main result of including scatter is that the upper bound on the gas density {\em become more stringent}. Additionally, the lower bound on metallicity tightens slightly. This results because, owing to the impact of collisional de-excitations, the ensemble-averaged luminosity of Eq \ref{eq:l10_avg} is smaller than the luminosity at the mean gas density. In this sense, our bounds derived ignoring scatter are conservative. The results are further quantified in Table \ref{tb:2}: the typical shift in the $1-\sigma$ gas density constraint is 0.7 dex, while the metallicity bound shifts by about 0.08 dex. 

Note that in comparing with the ALMA data we assumed our fiducial \textsc{starburst99} ionizing spectrum. A softer ionizing spectrum with $T_{4,\mathrm{eff}} \lesssim 4$ would lead to a noticeably smaller $V_{\mathrm{OIII}}/V_{\mathrm{HII}}$ ratio (Fig \ref{fig:V_correction}), which would demand larger metallicities and/or smaller gas densities. This case might be hard to reconcile with the ALMA data, or at least with the galaxies with large $L_{10}/\mathrm{SFR} \sim 10^{7.5} L_\odot/M_\odot/\mathrm{yr}$. Our model has also assumed $1-f_{\mathrm{esc}}=1$; if an appreciable fraction of ionizing photons escape into the IGM, then our constraint would tighten as $L_{10} \propto (1-f_{\mathrm{esc}})$ in our model. 

Finally, the results of Table \ref{tb:2} assume that the stellar metallicity matches the gas metallicity. However, as mentioned earlier, the spectra of $z \sim 2-3$ LBGs may be best explained if the stellar metallicity is as much as a factor of $\sim 5$ smaller than the gas-phase metallicity. In this case, since the $Q_{\mathrm{HI}}-\mathrm{SFR}$ relation depends on stellar metallicity (Eq \ref{eq:q_sfr}), our fiducial assumption may underestimate $Q_{\mathrm{HI}}$ and $L_{10}$ for a given SFR. If we apply the same factor of $5$ suggested by \cite{Steidel:2016hvv} at $z \sim 2-3$ to the $z \sim 6-9$ ALMA sample, we find that this loosens the metallicity bound by $0.05-0.13$ dex and the gas density limit by $0.04-0.25$ dex. The precise shifts here depend on the how stringent the bounds are before accounting for the reduction in stellar metallicity, but are relatively modest given current error bars. 

In general, we find good agreement with the recent work of \cite{Jones20}, which also derives metallicity bounds for some of the same galaxies considered here. One difference with these authors is that they fix $n_e=250$ cm$^{-3}$ and $T=1.5 \times 10^4$ K, while variations around these parameters are considered as part of a systematic error budget.\footnote{For further comparison, we can compare with their Eq 2 at a typical example metallicity of $Z=0.2 Z_\odot$. This equation gives their version of the $L_{10}-\mathrm{SFR}$ relation derived, in their case, from the \textsc{PyNeb} \citep{2015A&A...573A..42L} code. Here we ignore the ionization correction, since we find this to be less important than the 0.17 dex fiducial, empirically-motivated value from \cite{Jones20}, which is, in any case, small compared to their systematic error budget of 0.4 dex. Here our $L_{10}-\mathrm{SFR}$ relation agrees with these authors to within 0.03 dex at $Z=0.2 Z_\odot$, $n_{\mathrm{e}}=250$ cm$^{-3}$, and $T=1.5 \times 10^4$ K, illustrating excellent agreement between our independent modeling efforts.} The range in gas density considered is $n_e=10-600$ cm$^{-3}$ based on limits inferred from the MOSDEF survey near $z \sim 2$ \citep{Sanders16}, while we remain more agnostic about the density in this work (our prior spans $0\leq\log (n_{\mathrm{H}}/\mathrm{[cm^{-3}]})\leq4.0$).
Although their methodology is different than ours, 
for most of the overlapping sample considered, our $1-\sigma$ lower bound on metallicity agrees with their $1-\sigma$ lower metallicity limits to within $0.4$ dex. This difference is consistent with their systematic error budget, which is estimated at $0.4$ dex.
The one exception is Y1, for which our lower bound is $0.7$ dex  smaller than their limit; this may be related to the treatment of the very asymmetric error bar in this galaxy or it may owe to differences in our methodologies.  As discussed in \cite{Jones20}, these constraints are also in broad agreement with earlier work from e.g. \cite{2016Sci...352.1559I}, which compared NB1006-2 with \textsc{CLOUDY} models of [\oiii] 88$\mu$m emission. 

\par

The broad implication of these results is that the ISM is highly enriched with heavy elements in at least some of the current sample of ALMA galaxies; this occurs within only $\sim 500$ Myr to $1$ Gyr after the big bang. For example, $Z \gtrsim 0.6 Z_\odot$ in J0217 at $z=6.2$, when the age of this universe is $880$ Myr. \cite{Jones20} estimate that many of the ALMA galaxies formed stars for $\sim 100$ Myr prior to $z \sim 7$.  This is in general agreement with earlier spectral energy distribution modeling which found evidence for evolved stellar populations in some of the [\oiii] emitting galaxies (e.g. \citealt{2018Natur.557..392H}).

\section{Additional [\oiii] Lines}\label{sec:diagnostics}

As also emphasized in \cite{2018MNRAS.481L..84M} and \cite{Jones20}, additional [\oiii] line transitions should be observable with ALMA and the JWST. These can be fruitfully combined with star-formation rate indicators, such as hydrogen recombination line measurements. 
The SPHEREx project \citep{2014arXiv1412.4872D}
will also provide line-intensity mapping observations of some of the [\oiii] rest-frame optical transitions. In the line-intensity mapping context, observing multiple different transitions from gas at a given redshift is invaluable for mitigating foreground interloper contamination and other systematics \citep{Lidz:2016lub}. In terms of scientific yield, probing multiple transitions can help in empirically constraining the electron density, temperature, and ionization state of the gas. Here we briefly discuss modeling additional [\oiii] lines and their diagnostic power. We refer the reader to \cite{2018MNRAS.481L..84M} and \cite{Jones20} for a discussion regarding the prospects for observing these additional lines with ALMA and the JWST. 

\subsection{[\oiii] 52 $\mu$m line}\label{sec:5.1}
Our model can be easily extended to the case of the [\oiii] 52 $\mu$m fine structure line emitted in the $^3\mathrm{P}_2\rightarrow^3\mathrm{P}_1$ transition. Specifically, following the derivation of Eq \ref{eq:l10_general} the luminosity in this line is:
\begin{equation}\label{eq:18}
\begin{split}
    L_{21}&=n_2\dfrac{A_{21}}{1+0.5\tau_{21}}h\nu_{21}V_{\oiii}\\
    &=\dfrac{R_2}{1+R_1+R_2}\left(\dfrac{n_{\mathrm{O}}}{n_{\mathrm{H}}}\right)_\odot\dfrac{Z}{Z_\odot}\dfrac{A_{21}}{1+0.5\tau_{21}}h\nu_{21}\dfrac{Q_{\mathrm{HI}}}{\alpha_{\mathrm{B,HII}}n_e}\dfrac{V_{\mathrm{OIII}}}{V_\mathrm{HII}}\,.
\end{split}
\end{equation}
 Similar to the $L_{10}$ case, we find that Eq \ref{eq:18} agrees with the CLOUDY simulations to within 15\% fractional error over the full range of ISM and population synthesis models considered in this work. 
 
\begin{figure*}
    \centering
    \includegraphics[width=1.0\textwidth]{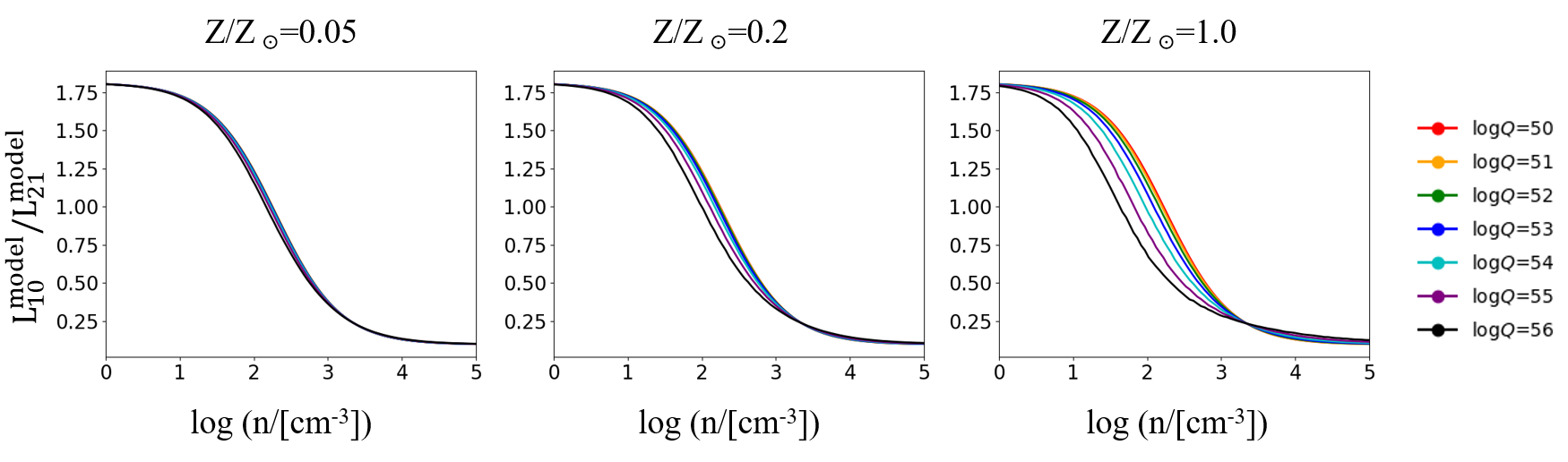}
    \caption{Ratio between the luminosity of [\oiii] 88 $\mu$m and [\oiii] 52 $\mu$m emission lines, as a function of gas density, computed using Eq (\ref{eq:21}) at $T=10^4$ K. The three panels show cases with varying gas metallicity ($Z=0.05 Z_\odot$, $Z=0.2 Z_\odot$, and $Z=Z_\odot$ from left to right), while the different colored lines show different ionization rates, $Q_{\mathrm{HI}}$. Each model assumes our fiducial \textsc{starburst99} ionizing spectrum. }\label{fig:l10_l21}
\end{figure*}

As a consequence of the differing critical densities between the two emission lines ($n_{e,\mathrm{crit}}=1732$ cm$^{-3}$ for $L_{10}$ and $n_{e,\mathrm{crit}}=A_{21}/k_{21}=4617$ cm$^{-3}$ for $L_{21}$ at $T=10^4$ K), the line ratio $L_{10}/L_{21}$ behaves roughly as a step-function in density, $n_e$ (e.g. \citealt{2011piim.book.....D}). To see this, note that in the low density limit of each line:
\begin{equation}
    \dfrac{n_1}{n_2} \approx \dfrac{k_{01}+k_{02}}{A_{10}}\Bigg/\dfrac{k_{02}}{A_{21}}=\dfrac{k_{01}+k_{02}}{k_{02}}\dfrac{A_{21}}{A_{10}}\,.
\end{equation}
In the opposite, high density limit, collisional de-excitations dominate over spontaneous decays and the level populations are determined by the Boltzmann factor: $\dfrac{n_u}{n_l}=\dfrac{g_u}{g_l}e^{-E_{ul}/(kT)}\approx \dfrac{g_u}{g_l}$, where $u$ and $l$ denote upper and lower levels. The relative occupancy is weakly dependent on temperature because the energy separation between the levels  here is small relative to the temperature of the HII region gas (see Fig \ref{fig:oiii_levels}). Therefore, the luminosity ratio in the two lines in the respective asymptotic limits is:
\begin{equation}\label{eq:21}
\dfrac{L_{10}}{L_{21}}\\
=\begin{cases}
\dfrac{k_{01}+k_{02}}{k_{02}}\dfrac{\nu_{10}}{\nu_{21}}=1.8 & \text{if $n_e\rightarrow0$}; \\
\\
\dfrac{g_1}{g_2}\dfrac{A_{10}}{A_{21}}\dfrac{\nu_{10}}{\nu_{21}}=0.1 & \text{if $n_e\rightarrow\infty$},
\end{cases}
\end{equation}
where the numerical values assume $T_4=1$. At intermediate densities, the line ratio smoothly interpolates between these two limits (see Fig \ref{fig:l10_l21}). Note that the line ratio depends weakly on metallicity and $Q_{\mathrm{HI}}$, with the small dependence arising from the impact of radiative trapping, as the escape probability decreases with increasing Str$\ddot{\mathrm{o}}$mgren radius and metallicity. For $v_{\mathrm{turb}} \sim 100$ km/s, radiative trapping is only a small effect, however. 
The figure illustrates that this line ratio provides a density diagnostic: specifically, one can robustly determine whether measurements are to the low density side or high density side of the step near the critical density. This can help in breaking degeneracies between the metallicity and the gas density that impact the constraints from a single [\oiii] line (e.g. as in \S \ref{sec:ALMA_constraints}). Note also that the $52 \mu$m line is, in fact, more luminous at high densities than the $88 \mu$m one \citep{Jones20}.

\par

\subsection{[\oiii] optical lines}

We can further consider the [\oiii] rest-frame optical lines: we focus on the $^1\mathrm{D}_2\rightarrow  \, ^3\mathrm{P}_2$ transition at 5008 \AA\ and the $^1\mathrm{D}_2\rightarrow \, ^3\mathrm{P}_1$ line at 4960 \AA\, between states with level populations of $n_3 \rightarrow n_2$ and $n_3 \rightarrow n_1$, respectively. Here we exploit the fact that the energy gaps between these transitions are large relative to the gas temperature (see Fig \ref{fig:oiii_levels}), $k_B T$. Therefore, the ratio of the luminosity in these lines to that in the fine structure transitions provides a gas temperature diagnostic (e.g. \citealt{2011piim.book.....D}).  

\begin{figure}
    \centering
    \includegraphics[width=0.45\textwidth]{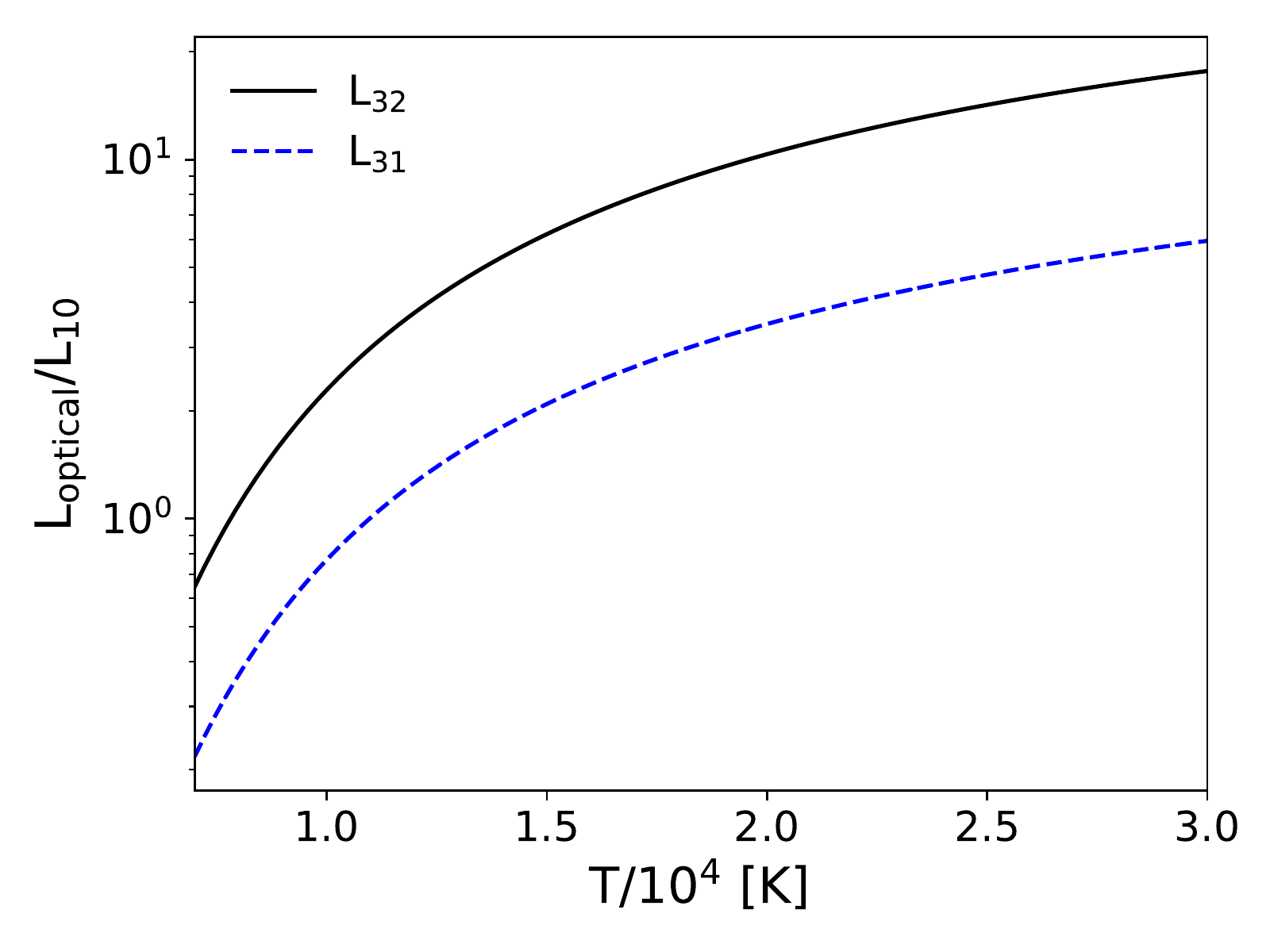}
    \caption{Temperature dependence of $L_{32}/L_{10}$ and $L_{31}/L_{10}$ predicted using Eq \ref{eq:27}. Here $L_{32}$ denotes the 5008 \AA\ transition, $L_{31}$ is the 4960 \AA\ line, and $L_{10}$ is the 88 $\mu$m fine-structure transition. }\label{fig:temp_diagnostic}
\end{figure}

For simplicity, we work here in the low density limit of the fine-structure lines; note also that the critical densities in the optical lines are much larger. In this case, following an identical logic to that leading up to Eq \ref{eq:l10low_qhi},
the luminosity ratio between the [\oiii] 5008 (4960) \AA\ line and [\oiii] 88 $\mu$m line is:
\begin{equation}\label{eq:27}
\begin{split}
    \dfrac{L_{32}}{L_{10}}&=\dfrac{3}{4}\dfrac{k_{03}}{k_{01}+k_{02}}\dfrac{\nu_{32}}{\nu_{10}}\,,\\
    \\
    \dfrac{L_{31}}{L_{10}}&=\dfrac{1}{4}\dfrac{k_{03}}{k_{01}+k_{02}}\dfrac{\nu_{31}}{\nu_{10}}\,,
\end{split}
\end{equation}
where we have used that $A_{32}/A_{31} \approx 3$. The line ratio is independent of the gas metallicity, ionization correction, and the gas density (in the low-density limit assumed here). However, the ratio is sensitive to the gas temperature via the temperature dependence of the collisional excitation coefficients that enter. Therefore, it can help to constrain the gas temperature, as illustrated in Fig \ref{fig:temp_diagnostic}. We find that the luminosities $L_{32}$ and $L_{31}$ predicted by our model agree with \textsc{CLOUDY} simulations to within 25\% fractional error at $Z \leq 0.2 Z_\odot$. At higher metallicities, the agreement is less good because the line ratio is quite sensitive to the temperature and our simple empirical temperature calibration (Eq \ref{eq:3}) is less accurate at high metallicity.

\par

In addition to the [\oiii] lines considered here, rest frame optical [\oii] emission lines are a promising target (e.g. \citealt{Jones20}). Specifically, joint measurements of [\oii] and [\oiii] emission lines can be used to empirically pin-down the $V_{\mathrm{OIII}}/V_{\mathrm{HII}}$ correction and the ionization state of the gas. We defer, however, modeling [\oii] lines to future work. 

\section{Discussion and conclusion}\label{sec:conclusions}

We introduced an analytic model aimed at describing the relationship between a galaxy's [\oiii] 88 $\mu$m luminosity and its star-formation rate. The model treats the total emission of ionizing radiation across a galaxy as though it were concentrated in a single effective source of radiation at the center of a spherically symmetric ionized region. The volume of ionized hydrogen gas is determined assuming a steady-state Str$\ddot{\mathrm{o}}$mgren balance between photoionizations and recombinations. The [\oiii] emission is subsequently predicted according to a three-level atom description, accounting for collisional excitations, de-excitations, radiative trapping, and an ionization/volume correction to account for the relative volumes of doubly-ionized oxygen and hydrogen gas. The model assumes that each HII region has a uniform density, metallicity, temperature, and ionizing spectrum, although local variations in the gas density and metallicity are approximately treated by allowing lognormal scatter in these quantities.

In the low density limit, the [\oiii] luminosity is linearly proportional to the gas metallicity and independent of gas density. At higher density, the luminosity decreases owing to collisional de-excitations. After accounting for plausible levels of line broadening, the effects of radiative trapping are fairly unimportant. We cross-check our model against detailed \textsc{CLOUDY} simulations and find agreement to within roughly 10-15\% accuracy across a broad range of parameters. Although we focus here on high redshift science applications, we believe our calculations apply more broadly, down to $z \sim 0$. The only aspect of our model that is calibrated to \textsc{CLOUDY} rather than predicted from first principles is the volume-averaged gas temperature (Eq \ref{eq:3}). Our model helps to clarify the key physical processes involved in determining the [\oiii] emission signal and can be used to rapidly explore parameter space. 

Although this agreement is encouraging, it would be interesting to compare our results with numerical simulations capable of -- at least partly -- resolving relevant ISM properties across simulated galaxies (e.g. \citealt{Hopkins:2017ycn,2019MNRAS.487.5902K}). These simulations are necessary to capture the detailed geometry and dynamics of the ionized gas in high redshift galaxies, and cross-check our simplified treatment. Although our model is obviously incapable of capturing the full complexity of the [\oiii] emission across a galaxy, it should provide a more accurate description of the galaxy's total [\oiii] luminosity and its dependence on average ISM properties. In its current implementation, we consider galaxy-averaged properties but our approach could be modified to describe the emission from individual HII regions. It would also be interesting to compare this work with simulations of individual HII regions (e.g. \citealt{2011MNRAS.414.1747A,2014MNRAS.445.1797M}), which can help test our simplified geometry, neglect of dynamical effects and magnetic fields, and other simplifications.\par
 
We use our model to extract constraints on ISM parameters from the current sample of nine $z \sim 6-9$ ALMA compiled in \cite{Harikane19}. Specifically, we use the observed [\oiii] 88 $\mu$m luminosity to star-formation rate ratio to derive constraints on the gas metallicity and density. The current data sample shows a fairly large scatter in the [\oiii] 88 $\mu$m luminosity to star-formation rate ratio, suggesting a diversity in ISM properties at $z \sim 6-9$. We report lower bounds on gas metallicity and upper bounds on gas density, which vary across the current sample by as much as 1.8 dex. The tightest bounds in the sample give $Z \gtrsim 0.5 Z_\odot$ and $n_{\mathrm{H}} \lesssim 50$ cm$^{-3}$ at around $2-\sigma$ confidence, indicating a remarkable level of chemical enrichment in some systems within the first Gyr after the big bang. 
In this comparison, we fixed the ionizing spectral shape to that of a \textsc{starburst99} continuous star-formation rate model with an age of 10 Myr. We believe that the data prefer a relatively hard spectrum (with the 1 to 4 Rydberg spectrum being harder than a blackbody with temperature $T_{\mathrm{4,eff}} \geq 4$, see \S \ref{sec:voiii_corr}), so that the volume of doubly ionized oxygen gas is comparable to that of ionized hydrogen. It might be interesting to compare with a wider suite of population synthesis models including binarity (e.g. \citealt{Stanway14}) to better understand the implications of these results for stellar population models. 

An additional caveat in our analysis is that we ignore the presence of dust grains, which can absorb hydrogen ionizing photons and limit the volume of ionized hydrogen and doubly-ionized oxygen gas in the galaxy. Note, however, that including this effect would make our bounds on metallicity and gas density {\em more stringent}. Another limitation with our current constraints comes from the degeneracy between metallicity and gas density.
Empirically, this degeneracy may be partly broken using future ALMA measurements of the 52 micron [\oiii] emission line. Further, measurements of [\oiii] and [\oii] optical transitions can help constrain the temperature and ionization state of the emitting gas.

Our model may also be combined with numerical simulations and/or semi-analytic models of galaxy formation. In particular, these models can be used to predict plausible distributions in gas metallicity, density, and any correlations between these properties and stellar mass, star-formation rate, or host host halo mass. In the future, additional [\oiii] measurements across a range of redshifts may help to determine the chemical enrichment history of galaxies and provide a powerful test of simulations of galaxy formation and evolution. 

We also intend to extend our model to describe line-intensity mapping signals since [\oiii] is a promising target line for such studies. One approach for describing the line-intensity mapping signal is to parameterize the mass-metallicity relationship, while further connecting stellar mass, star-formation rate, and host halo mass via abundance matching. This effort will connect the line-intensity mapping signal to interesting ISM properties; notably, line-intensity mapping efforts are uniquely powerful in determining aggregate properties, including the impact of galaxies that are too faint to detect directly. These studies may therefore provide a valuable complement to the ALMA measurements. While we focus on [\oiii] in this work, it will be interesting to generalize our analysis to consider further emission lines. The ultimate power of line-intensity mapping, as well as targeted measurements towards individual galaxies, naturally comes from detecting as many emission lines as possible. 

\section{Acknowledgement}
We acknowledge Yuichi Harikane for providing sample \textsc{CLOUDY} input files and Christophe Morisset for discussions regarding \textsc{CLOUDY}. We thank Hamsa Padmanabhan for useful comments on a draft manuscript, and Patrick Breysse and Lile Wang for helpful conversations. 
AL acknowledges support through NASA ATP grant 80NSSC20K0497. 

\section*{Data availability}
The data used to support the findings of this study are available from the corresponding author upon request.




\bibliographystyle{mnras}
\bibliography{LOIII}




\label{lastpage}
\end{document}